%% file: Article.tex
\documentclass[11.5pt]{article}

\usepackage[utf8]{inputenc}
\usepackage[margin=1.0in]{geometry}

\usepackage{upgreek}
\usepackage{graphicx}
\usepackage{float}
\graphicspath{ {Images/} }

\usepackage{cite}
\usepackage{braket}
\usepackage{amsmath}
\usepackage{gensymb}
\usepackage{multicol}
\usepackage{authblk}

\title{Understanding the Methyl-TROSY effect over a wide range of magnetic fields}
\author[1]{Nicolas Bolik-Coulon}
\author[1]{Samuel F.Cousin}
\author[1]{Pavel Kade$\mathrm{\check{r}\acute{a}}$vek}
\author[2]{Jean-Nicolas Dumez}
\author[1]{Fabien Ferrage}

\affil[1]{Laboratoire des Biomolécules, LBM, Département de Chimie, École Normale Supérieure, PSL University, Sorbonne Université, CNRS, 75005 Paris, France}
\affil[2]{CEISAM, CNRS, Université de Nantes, 44300 Nantes, France}

\begin{document}
\maketitle
\input{NewCommand.tex}

\begin{abstract}
\input{Section/Abstract}
\end{abstract}

\input{Section/1_Introduction}

\input{Section/2_Methods}

\input{Section/3_TheoreticalFramework}

\input{Section/4_HighField}

\input{Section/5_Generalization}

\input{Section/6_ComparisonExp}

\input{Section/7_Conclusion.tex}

\section{Acknowledgment}
The authors thank Malcolm Levitt (University of Southampton) for fruitful discussions. We thank Geoffrey Bodenhausen (\'Ecole Normale Sup\'erieure, Paris), Vitali Tugarinov (NIH, Bethesda) and Tairan Yuwen (University of Toronto) for their helpful comments of the preprint version of this manuscript. This project received funding from the European Research Council (ERC) under the European Union’s Seventh Framework Programme (FP7/2007-2013), ERC Grant agreement 279519 (2F4BIODYN).

\bibliographystyle{ieeetr}
\bibliography{Bibliography}

\section*{Supplementary Material}
\input{Section/SI.tex}

\end{document}

%% file: NewCommand.tex
\newcommand{\ham}{$\mathcal{H}$}
\newcommand{\HAM}{\mathcal{H}}
\newcommand{\CyHy}{2\hat{\mathrm{C}}_{\mathrm{y}}\hat{\mathrm{H}}_{\mathrm{y}}}
\newcommand{\OMc}{\omega_{_{\mathrm{C}}}}
\newcommand{\omegaC}{\omega_{\mathrm{C}}}
\newcommand{\gammaH}{\gamma_\mathrm{H}}
\newcommand{\gammaC}{\gamma_\mathrm{C}}
\newcommand\Cz{\hat{C}$_{\mathrm{z}}$ }
\newcommand\Hz{\hat{H}$_{\mathrm{z}}$ }
\newcommand\CzHz{\^C$_{\mathrm{z}}$\^H$_{\mathrm{z}}$ }
\newcommand\NzHz{\^N$_{\mathrm{z}}$\^H$_{\mathrm{z}}$ }
\newcommand\CpHm{2C$^+$H$^-$}
\newcommand\CmHp{2C$^-$H$^+$}
\newcommand\CpHp{2C$^+$H$^+$}
\newcommand\CmHm{2C$^-$H$^-$}
\newcommand{\ZQpm}{\mathrm{ZQ^{2\mathrm{C}^+\mathrm{H}^-}}}
\newcommand{\DQpp}{\mathrm{DQ^{2\mathrm{C}^+\mathrm{H}^+}}}

\newcommand{\Proton}{$^1$H}
\newcommand{\Carbon}{$^{13}$C}
\newcommand{\Nitrogen}{$^{15}$N}

\newcommand{\tauC}{\tau_{\mathrm{C}}}
\newcommand{\tauc}{$\uptau_{\mathrm{c}}$}
\newcommand{\taum}{$\uptau_{\mathrm{m}}$}
\newcommand{\TAUC}{\tau_c}
\newcommand{\TAUF}{\tau_f}
\newcommand{\TAUM}{\tau_{\mathrm{met}}}
\newcommand{\TAUS}{\tau_s}
\newcommand{\TAUEFF}{\tau_{eff}}

\newcommand{\fcorr}{$\mathcal{C}$(t)}
\newcommand{\HdO}{H$_2$O}
\newcommand{\DdO}{D$_2$O}
\newcommand{\Runrho}{R$_{1\uprho}$}
\newcommand{\sHHH}{\mathrm{HHH}}
\newcommand{\sHCH}{\mathrm{HCH}}
\newcommand{\sCHH}{\mathrm{CHH}}
\newcommand{\sCH}{\mathrm{CH}}
\newcommand{\sHH}{\mathrm{HH}}
\newcommand{\sCHHH}{\mathrm{CHHH}}
\newcommand{\OMh}{\omega_{_{\mathrm{H}}}}

\newcommand{\jdens}{$\mathcal{J}(\omega)$}
\newcommand\Jdens{$\mathcal{J}(\omega)$}
\newcommand{\JDENS}{\mathcal{J}(\omega)}
\newcommand{\JCC}{\mathcal{J}_{\mathrm{CC}}}
\newcommand{\JCH}{\mathcal{J}_{\mathrm{CH}}}
\newcommand{\JCHH}{\mathcal{J}_{\mathrm{CHH}}}
\newcommand{\JCCH}{\mathcal{J}_{\mathrm{CCH}}}
\newcommand{\JCHHH}{\mathcal{J}_{\mathrm{CHHH}}}
\newcommand{\JHH}{\mathcal{J}_{\mathrm{HH}}}
\newcommand{\JCCHH}{\mathcal{J}_{\mathrm{CCHH}}}
\newcommand{\JHCH}{\mathcal{J}_{\mathrm{HCH}}}
\newcommand{\JHHH}{\mathcal{J}_{\mathrm{HHH}}}

\newcommand{\CDDH}{$\{^{13}$C$^1$H$^2$H$_2\}$}
\newcommand\CHH{$^{13}$C$^1$H$_3$ }
\newcommand{\ubiquitinRelax}{U-$[$$^2$H, $^{15}$N$]$, Ile-$\updelta_1$[$^{13}$C$^1$H$^2$H$_2$]-ubiquitin}
\newcommand{\ubiquitin}{U-[$^2$H, $^{15}$N], Ile-$\updelta_1$[$^{13}$CH$_3$]-ubiquitin}

\newcommand{\SlowTumbling}{slow-tumbling}

\newcommand{\beginsupplement}{%
        \setcounter{table}{0}
        \renewcommand{\thetable}{S\arabic{table}}%
        \setcounter{figure}{0}
        \renewcommand{\thefigure}{S\arabic{figure}}%
     }

%% file: Section/Abstract.tex
\textbf{Abstract:} The use of relaxation interference in the methyl Transverse Relaxation-Optimized SpectroscopY (TROSY) experiment has opened new avenues for the study of large proteins and protein assemblies in nuclear magnetic resonance. So far, the theoretical description of the methyl-TROSY experiment has been limited to the \SlowTumbling~approximation, which is correct for large proteins on high field spectrometers. In a recent paper, favorable relaxation interference was observed in the methyl groups of a small protein at a magnetic field as low as 0.33\,T, well outside the \SlowTumbling~regime. Here, we present a model to describe relaxation interference in methyl groups over a broad range of magnetic fields, not limited to the \SlowTumbling~regime. We predict that the type of multiple-quantum transitions that show favorable relaxation properties change with the magnetic field. Under the condition of fast methyl-group rotation, methyl-TROSY experiments can be recorded over the entire range of magnetic fields from a fraction of 1\,T up to 100\,T.

%% file: Section/1_Introduction.tex
\section{Introduction}
Nuclear Magnetic Resonance (NMR) spectroscopy is a versatile tool to study protein structure and protein motions.\cite{Cavanagh_ProteinNMR_2007, Palmer_ChemRev_2004, Mittermaier_science_2006, Charlier_ChemSocRev_2016} The progress of NMR over the last decades was made possible by the constant development of magnets with higher magnetic fields,\cite{Ardenkjaer_Angewandte_2015} the availability of more sensitive probes, especially cryogenic probes, as well as countless innovative methodological developments. A major breakthrough for the investigation of large biomolecules has been the introduction of Transverse Relaxation-Optimized SpectroscopY (TROSY).\cite{Pervushin_attenuated_1997} In the  \Nitrogen-\Proton~spin pair present in peptide bonds, the TROSY effect relies on the selection of a coherence with destructive interference between the chemical-shift-anisotropy (CSA) and dipole-dipole (DD) relaxation mechanisms,\cite{Shimizu_JCP_1964, Goldman_JMagRes_1984, Wimperis_MolPhys_1989} leading to a dramatic decrease of the transverse relaxation rate.\cite{Pervushin_attenuated_1997,Tugarinov_fourdimensional_2002} The improvement in resolution and sensitivity has made possible the study of large biomolecular systems up to about 1\,MDa.\cite{fiaux_nmr_2002} \\
Side-chain methyl groups are also excellent probes to investigate the structure and dynamics of proteins as they are well spread over protein sequences.\cite{hajduk_nmr-based_2000} They give information about protein-ligand or protein-protein binding pockets, as well as the hydrophobic core of proteins where they act as an entropy reservoir.\cite{DuBay_AccChemRes_2015,Frederick_Nature_2007} Their motions are sensitive to the surrounding environment making them good probes for dynamic properties at specific positions of the protein.\cite{Nicholson_dynamics_1992, Cousin_JACS_2018, mas_SciAdv_2018} \\
Relaxation interference also gives rise to a TROSY effect in methyl \CHH groups in macromolecules. Tugarinov \emph{et al.}\cite{tugarinov_cross-correlated_2003} described the Heteronuclear Multiple Quantum Coherence (HMQC) experiment and associated methyl-TROSY effect using two main assumptions: the \SlowTumbling~approximation and infinitely fast methyl-group rotation. The \SlowTumbling~approximation is suitable to describe relaxation properties in high molecular-weight proteins on high-field magnets (10 to 25\,T). The rotation of methyl groups can be considered infinitely fast since it is much faster than the slow global tumbling of large proteins. Under these assumptions, the contributions of all intra-methyl DD couplings for the relaxation of the central line of the triplet are exactly zero in an HMQC experiment.\cite{tugarinov_cross-correlated_2003, Sheppard_Progress_2010} This major discovery hand in hand with the development of schemes for protein \CHH labeling at specific positions \cite{tugarinov_isotope_2006, clore_chapter_2012,mas_specific_2013} opened new perspective to study high-molecular weight proteins with NMR\cite{rosenzweig_bringing_2014} as was shown by several studies of large molecular machines, such as the proteasome \cite{tugarinov_JACS_2007} or a 1 MDa-chaperone.\cite{mas_SciAdv_2018} \\
Is the methyl-TROSY effect preserved beyond the current range of high magnetic fields? We have recently recorded a two-field Heteronuclear Zero Quantum Coherence (2F-HZQC) experiment on \ubiquitin.\cite{cousin_recovering_2016, tugarinov_line_2004} This experiment combines detection at 14.1\,T and evolution of multiple-quantum coherences at 0.33\,T by using a 2-field spectrometer.\cite{cousin_recovering_2016, cousin_high-resolution_2016} Narrow linewidths were obtained in the indirect low-field dimension suggesting a TROSY effect at low field, well outside the \SlowTumbling~regime. Understanding this observation requires a formal description of methyl group relaxation in all motional regimes.\\
Here we propose a general analysis of the relaxation properties of zero- and double-quantum (ZQ and DQ) coherences in methyl groups, which goes beyond the main hypotheses of the original methyl-TROSY work: \SlowTumbling~and fast methyl rotation. A formal analysis shows that the free evolution of multiple-quantum coherences occurs in a subspace of the Liouville space of dimension 4. Slowly relaxing components are predicted to exist in this subspace for arbitrary values of the magnetic field. At high fields, two coherences relax slowly, which, as expected, correspond to the central transitions of the MQ triplet. Numerical calculations of relaxation rates show that the calculated linewidths using our approach are in good agreement with experimental linewidths at both high and low fields. At very high magnetic fields, which are expected to be accessible in the future, we find that the interference between the chemical shift anisotropy and the sum of dipolar couplings leads to favorable relaxation properties for zero-quantum coherences that correspond to one of the outer lines of the triplet. This work provides a theoretical foundation for the development of methyl-TROSY experiments outside of the current range of high-field biomolecular NMR: in low-field NMR,\cite{Blumich_ACIE_2017} possibly in combination with hyperpolarization methods, in two-field NMR,\cite{cousin_recovering_2016, cousin_high-resolution_2016} or at very high fields, already accessible with hybrid magnets.\cite{Gan_JMR_2017}

%% file: Section/2_Methods.tex
\section{Methods}
\subsection{NMR spectroscopy}
\begin{figure}
	\begin{center}
		\includegraphics[width=0.9\textwidth]{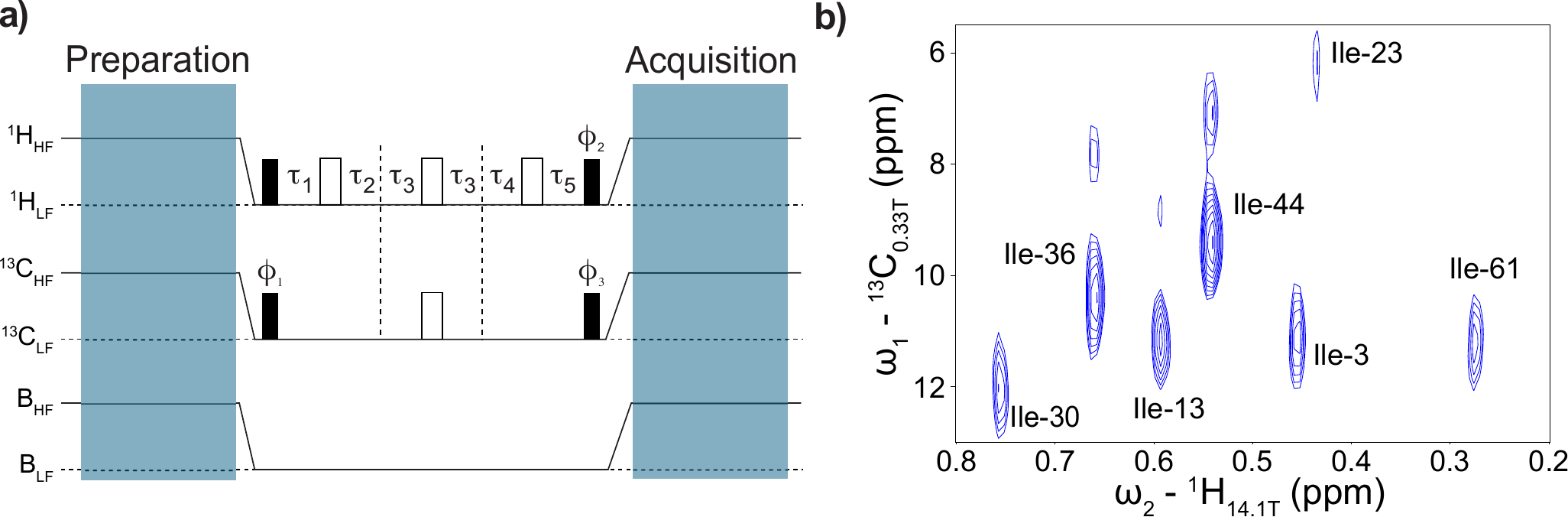}
	\end{center}
	\caption{Two-field HZQC experiment recorded on \ubiquitin. \textbf{a.} Schematic representation of the pulse sequence (see \cite{cousin_recovering_2016} for details). All pulses are applied along the x-axis unless specified otherwise. The phases are cycled as $\phi_1 = x, -x$, $\phi_2 = 4\{x\}, 4\{y\}$ and $\phi_3 = 2\{x\}, 2\{-x\}, 2\{y\}, 2\{-y\}$. The phase cycling ensures that the spin system evolves under ZQ coherences during $\tau_1 = (\tau_0 + n_1 \Delta t_1) c$ and $\tau_5 = \tau_0 c$ and under DQ coherences during $\tau_2 = (\tau_0 + n_1 \Delta t_1)(1 - c)$, $2\tau_3 = 2\tau_0 (2c-1)+2n_1 \Delta t_1 (c - 0.5)$ and $\tau_4 = \tau_0 (1 - c)$ with $c = (\gamma_C/\gamma_H +1)/2$, $\Delta t_1$ the time increment and $n_1$ the index of the time increment. HF: High Field. LF: Low Field. \textbf{b.} 2D-correlation spectrum of the seven isoleucines recorded on \ubiquitin.\cite{cousin_recovering_2016} A shearing transformation allows the display of pure \Carbon~chemical shifts in the indirect dimension.}
	\label{fig:SpectrumPP}
\end{figure}
Experiments were performed on a 600 MHz Avance III NMR spectrometer, equipped with a two-field system apparatus\cite{cousin_high-resolution_2016} shuttling the sample between 14.1\,T and 0.33\,T. The high-field center is equipped with a TXI (\Proton, \Carbon, \Nitrogen) probe. The average shuttling time per transfer is 120\,ms. The low-field part of the two-field experiment which will be simulated later is shown in Fig\,\ref{fig:SpectrumPP}.a. The details of the 2F-HZQC experiment can be found in \cite{cousin_recovering_2016}. Experiments were performed on 1.5\,mM specifically labeled \ubiquitin~at pH 4.5 in 50\,mM AcAc (expressed and purified as detailed in \cite{cousin_recovering_2016}). The 2F-HZQC sepctrum is shown in Fig.\,\ref{fig:SpectrumPP}.b.

\subsection{Relaxation theory}
Analytical and numerical calculations have been performed using Mathematica\cite{wolfram_research_inc._mathematica_2016} with SpinDynamica package\cite{Bengs_MagnResonChecm_2017} to generate the appropriate operators, and taking advantage of the analytical approach proposed by I.Kuprov \emph{et al.} to implement the BRW theory.\cite{kuprov_bloch-redfield-wangsness_2007} The geometry of the methyl group was assumed to be tetrahedral, with the C-H bond length r$_{\mathrm{CH}}=1.105$\,\AA~and the angle sustended by the C-C and C-H bonds $\alpha\,=\,\cos^{-1}(1/3)\,=\,109.47\,\degree$.

\subsection{Simulation of the HZQC spectrum}
\paragraph{Relaxation}
The propagator for an evolution of the spin system during the period $\tau = \tau_1 + \tau_2 + 2\tau_3 + \tau_4 + \tau_5$, times $
\tau_i$ defined in Fig.\,\ref{fig:SpectrumPP}.a, at low field is (without considering the chemical shift evolution):
\begin{equation}
\begin{aligned}
	\mathcal{P}(\tau) =& e^{- \tau/4 \left(\hat{\hat{\mathcal{R}}}_{ZQ} +\hat{\hat{\mathcal{H}}}_J \right)} \mathcal{F}_{ZD}^{-1} e^{-\tau/4 \left(\hat{\hat{\mathcal{R}}}_{DQ} + \hat{\hat{\mathcal{H}}}_J \right) } \times  \mathcal{U} e^{-\tau/4 \left(\hat{\hat{\mathcal{R}}}_{DQ} + \hat{\hat{\mathcal{H}}}_J \right) } \mathcal{F}_{ZD} e^{-\tau/4 \left(\hat{\hat{\mathcal{R}}}_{ZQ}  + \hat{\hat{\mathcal{H}}}_J \right)}
    \label{eq:Propagator}
\end{aligned}
\end{equation}
where $\hat{\hat{\mathcal{R}}}_{ZQ}$ (resp. $\hat{\hat{\mathcal{R}}}_{DQ}$) is the relaxation matrix superoperator in the zero- and double-quantum operator bases $\mathcal{B}_{ZQ}$ (resp. $\mathcal{B}_{DQ}$) basis (\textit{vide infra}), $\mathcal{U}$ the matrix accounting for the effect of the simultaneous $\pi$-pulses:
\begin{equation}
\mathcal{U} = 
	\begin{bmatrix}
		\begin{array}{@{}*{4}{c}@{}}
			-1 & 0 & 0 & 0 \\
			0 & -1 & 0 & 0 \\
			0 & 0 & 0 &-1 \\
			0 & 0 & -1 & 0
		\end{array}
	\end{bmatrix}
\end{equation}
in basis \{$MQ_\mathrm{central}^{\Sigma \mathrm{E}}$, $MQ_\mathrm{central}^{\mathrm{A}}$, $MQ_\mathrm{outer,1}^{\mathrm{A}}$, $MQ_\mathrm{outer,2}^{\mathrm{A}}$\}, $\hat{\hat{\mathcal{H}}}_J$ accounts for the scalar coupling:
\begin{equation}
\hat{\hat{\mathcal{H}}}_J = 
	\begin{bmatrix}
		\begin{array}{@{}*{4}{c}@{}}
			0 & 0 & 0 & 0 \\
			0 & 0 & 0 & 0 \\
			0 & 0 & 2i\pi J & 0\\
			0 & 0 & 0 & -2i\pi J
		\end{array}
	\end{bmatrix}
\end{equation}
with \textit{J} the scalar coupling constant, set to 130\,Hz, and $\mathcal{F}_{ZD}$ the bijection function from $\mathcal{B}_{ZQ}$ to $\mathcal{B}_{DQ}$ accounting for the first and third proton $\pi$-pulses:
\begin{equation}
\mathcal{F}_{ZD} 
	\begin{bmatrix}
		\begin{array}{@{}*{1}{c}@{}}
			ZQ_\mathrm{central}^{\Sigma \mathrm{E}} \\
			ZQ_\mathrm{central}^{\mathrm{A}} \\
			ZQ_\mathrm{outer,1}^{\mathrm{A}} \\
			ZQ_\mathrm{outer,2}^{\mathrm{A}}
		\end{array}
	\end{bmatrix}
= -
	\begin{bmatrix}
		\begin{array}{@{}*{1}{c}@{}}
			DQ_\mathrm{central}^{\Sigma \mathrm{E}} \\
			DQ_\mathrm{central}^{\mathrm{A}} \\
			DQ_\mathrm{outer,2}^{\mathrm{A}} \\
			DQ_\mathrm{outer,1}^{\mathrm{A}}
		\end{array}
	\end{bmatrix}
\end{equation}
The expected value at the end of the low field evolution is obtained by:
\begin{equation}
	\mathcal{E}_{2\hat{C}_y\hat{H}_y} (\tau) = \rho_{2\hat{C}_y\hat{H}_y \rightarrow \mathcal{B}_{ZQ}} \mathcal{P}(\tau)\rho^{\mathrm{T}}_{2\hat{C}_y\hat{H}_y \rightarrow \mathcal{B}_{ZQ}}
    \label{eq:Propagator}
\end{equation}
where $\rho_{2\hat{C}_y\hat{H}_y \rightarrow \mathcal{B}_{ZQ}}$ is the projection of the operator monitored during the evolution period $2\hat{C}_y\hat{H}_y$ on $\mathcal{B}_{ZQ}$ and the superscript T refers to the transpose operation:
\begin{equation}
\begin{aligned}
\rho_{2\hat{C}_y\hat{H}_y \rightarrow \mathcal{B}_{ZQ}} &= \left[\frac{1}{2\sqrt{6}}, \frac{1}{2\sqrt{3}}, \frac{1}{4}, \frac{1}{4}\right]\\
\rho^{\mathrm{T}}_{2\hat{C}_y\hat{H}_y \rightarrow \mathcal{B}_{ZQ}} &=
	\begin{bmatrix}
		\begin{array}{@{}*{1}{c}@{}}
			1/(2\sqrt{6}) \\
			1/(2\sqrt{3}) \\
			1/4 \\
			1/4
		\end{array}
	\end{bmatrix}
\end{aligned}
\end{equation}
\paragraph{Detection} The entire spectrum was simulated by including the chemical shift evolution, with the same sweep width and number of points as in the recorded 2F-HZQC to provide a reliable comparison. Both simulated and measured spectra were processed with the same parameters for zero-filling. No apodisation function was used in the indirect dimension. The individual free induction decays were simulated and fast Fourier transformation was applied using the Python scipy.fftpack library. Spectra were created using the Python nmrglue library.\cite{Helmus_JbiolNMR_2013} All extracted slices were further normalized independantely to the maximum peak intensity.

%% file: Section/3_TheoreticalFramework.tex
\section{Theoretical framework}
\begin{figure*}
	\begin{center}
		\includegraphics[width=1.0\textwidth]{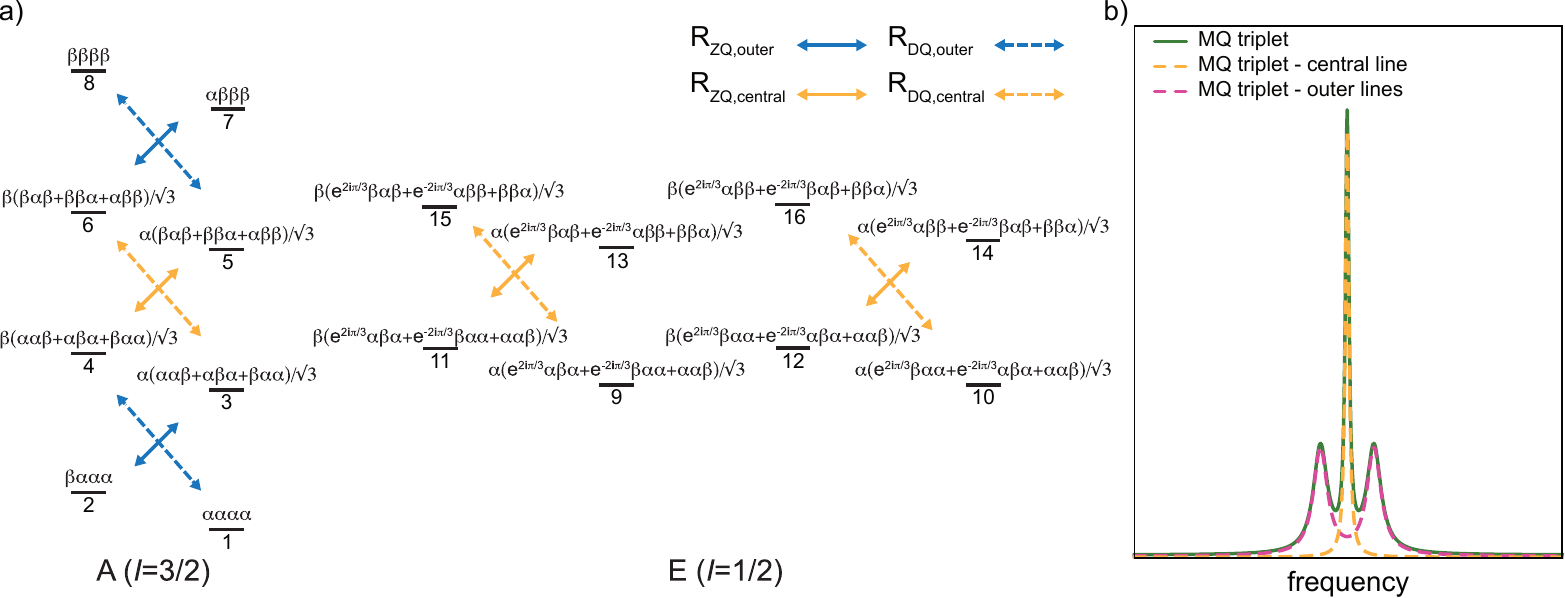}
	\end{center}
	\caption{Description of the methyl-TROSY multiple-quantum spectrum and associated spin transitions. \textbf{a.} Energy level diagram of a $^{13}$C$^1$H$_3$ spin system using symmetry-adapted states for the point group $C_3$. In the state labels, the first spin state corresponds to the \Carbon~nuclear spin, and the others to the three \Proton~spins \textit{i.e.} each spin state can be written as $\ket{\mathrm{C}\mathrm{H}_1\mathrm{H}_2\mathrm{H}_3}$ where C and $\mathrm{H}_i$ refer to the spin states of the carbon and proton \textit{i} respectively (either $\alpha$ or $\beta$). Each state is associated with a number in order to simplify the description. Transitions of interest have been highlighted with solid (resp. dashed) arrows for the ZQ (resp. DQ) coherences. Transitions giving rise to the outer (resp. central) component of the triplet are colored in blue (resp. orange). \textbf{b.} Simulated multiple-quantum methyl-TROSY triplet showing the contributions from the outer and central lines.}
	\label{fig:BasisMethyl}
\end{figure*}

\subsection{Spin system}
Many studies have focussed on the relaxation properties of an H$_3$ and $^{13}$C$^1$H$_3$ spin systems.\cite{Werbelow_JCP_1975, matson_methyl_1976, matson_JCP_1977, Werbelow_Advances_1977, Muller_JMR_1987, Kay_JMR_1992, Dumez_JCP_2015} Here, we consider a $^{13}$C$^1$H$_3$ spin system of an isolated methyl group for which tetrahedral geometry is assumed, that is the carbon nucleus occupies the center of a tetrahedron, three corners of which are occupied by the protons. This nuclear spin system is characterized by sixteen energy levels, direct product of the spin states of a \Carbon~spin with a \Proton$_3$ spin system. Mathematically, the relevant symmetry group to describe this system of three protons is the alternating group $A_3$.\cite{LevittC3} The point group $C_3$ is isomorphic to $A_3$ and more commonly used when refering to symmetry properties in physical sciences. An equivalent theoretical description could be built on the irreducible representation of the point group $C_{3v}$. The energy levels are then separated into two manifolds based on their spin quantum number: A ($I=\frac{3}{2}$) and E ($I=\frac{1}{2}$). A schematic representation is shown in Fig.\,\ref{fig:BasisMethyl}.a where the Zero Quantum (ZQ) and Double Quantum (DQ) coherences are highlighted in solid and dashed arrows respectively. \\

\subsection{Operators}
We will use the following convention: in the abscence of indices to \Proton~operators, the sum over the three protons is implicit, \textit{i.e.} $\hat{\mathrm{H}}^\pm= \sum_{i=1}^3 \hat{\mathrm{H}}^\pm_i/\sqrt{3}$ where $\hat{\mathrm{H}}_i$ is the operator for proton \textit{i}. The product operators that are relevant to the present analysis are the ZQ (\CpHm $\pm$ \CmHp) and DQ (\CpHp $\pm$ \CmHm) coherence operators of a $^{13}$C$^1$H$_3$ spin system. The analysis performed here only considers the \CpHm~and \CpHp~transition operators, but can be performed similarly for the corresponding \CmHp~and \CmHm~transition operators. A symmetry analysis shows that the subspace of operators for which: i) the point group is $C_3$ (i.e., the operators are unchanged by a circular permutation of the three protons) and ii) the coherence orders are m$_\mathrm{C}$ = +1 and m$_\mathrm{H}$ = -1 for ZQ transitions, and m$_\mathrm{C}$ = +1 and m$_\mathrm{H}$ = +1 for DQ transitions has dimension five. All transitions are shown in Fig.\,\ref{fig:BasisMethyl}.a. As explained below, defining the sums and the differences of the transitions in the E-manifold allows a size-reduction of the basis to four terms. A suitable basis for the study of the considered single-transition operators expressed in terms of the transitions shown in Fig.\,\ref{fig:BasisMethyl}.a is presented in Table\,\ref{Table:LabelOp}.

\begin{table}
	\caption{Single-transition operator basis used for the study of the relaxation properties of ZQ and DQ coherences. The numbers in the bra and ket refer to the states as shown in Fig.\,\ref{fig:BasisMethyl}.}%
	\begin{center}
		{\def\arraystretch{1.3}
		\begin{tabular}{cc|cc}
			\multicolumn{2}{c|}{\textbf{$\mathcal{B}_{ZQ}$}} & \multicolumn{2}{c}{\textbf{$\mathcal{B}_{DQ}$}} \\
			\hline
		 	$ZQ_{\mathrm{outer,1}}^\mathrm{A}$ & $\ket{3}\bra{2}$  & $DQ_{\mathrm{outer, 1}}^\mathrm{A}$ &  $\ket{4}\bra{1}$ \\%
		 	$ZQ_{\mathrm{outer,2}}^\mathrm{A}$ &  $\ket{7}\bra{6}$  & $DQ_{\mathrm{outer, 2}}^\mathrm{A}$ &   $\ket{8}\bra{5}$   \\%
		 	$ZQ_{\mathrm{central}}^\mathrm{A}$ &  $\ket{5}\bra{4}$  & $DQ_{\mathrm{central}}^\mathrm{A}$ &   $\ket{6}\bra{3}$   \\%
		 	$ZQ_{\mathrm{central}}^\mathrm{\Sigma E}$ &  $\frac{1}{\sqrt{2}}(\ket{13}\bra{11}$  & $DQ_{\mathrm{central}}^\mathrm{\Sigma E}$ &   $\frac{1}{\sqrt{2}}(\ket{15}\bra{9}$   \\%
			& $+\ket{14}\bra{12})$ & & $+\ket{16}\bra{10})$ 
		\end{tabular}
		}
	\end{center}
	\label{Table:LabelOp}
\end{table}

The expansion of the two considered multiple-quantum transition operators in the respective bases defined in Table\,\ref{Table:LabelOp} is:
\begin{equation}
	\begin{aligned}
   		2\mathrm{C}^{+}\mathrm{H}^{-} =& \frac{1}{4} ZQ_{\mathrm{outer,1}}^\mathrm{A} + \frac{1}{4} ZQ_{\mathrm{outer,2}}^\mathrm{A}  + \frac{1}{2\sqrt{3}} ZQ_{\mathrm{central}}^\mathrm{A} + \frac{1}{2\sqrt{6}} ZQ_{\mathrm{central}}^\mathrm{\Sigma E} \\ 
    		2\mathrm{C}^{+}\mathrm{H}^{+} =& \frac{1}{4} DQ_{\mathrm{outer, 1}}^\mathrm{A} + \frac{1}{4} DQ_{\mathrm{outer, 2}}^\mathrm{A}  + \frac{1}{2\sqrt{3}} DQ_{\mathrm{central}}^\mathrm{A} + \frac{1}{2\sqrt{6}} DQ_{\mathrm{central}}^\mathrm{\Sigma E}
    \end{aligned}
   \label{eq:MQdecomp}
\end{equation}
The single-transition operators $ZQ^\mathrm{\Delta E} = (\ket{13}\bra{11}-\ket{14}\bra{12})/\sqrt{2}$ and $DQ^\mathrm{\Delta E} = (\ket{15}\bra{9}-\ket{16}\bra{10})/\sqrt{2}$ need not to be included in the basis, as they are found to be in independent subspaces for the evolution analysed here: they do not contribute to the expansion of the studied ZQ and DQ transitions and they do not cross-relax with the other terms.\\
In the \SlowTumbling~approximation, the methyl-TROSY spectrum can be studied using two single-transition operators, equivalent to those described by Tugarinov \emph{et al.} \cite{tugarinov_cross-correlated_2003} for the description of the central and outer lines of the triplet (Fig.\,\ref{fig:BasisMethyl}.b):
\begin{equation}
	\begin{aligned}
   		T_{\mathrm{outer}} =& \frac{1}{4} \left(ZQ_{\mathrm{outer,1}}^\mathrm{A} + DQ_{\mathrm{outer,1}}^\mathrm{A}  + ZQ_{\mathrm{outer,2}}^\mathrm{A} + DQ_{\mathrm{outer,2}}^\mathrm{A} \right) \\ 
    		T_{\mathrm{central}} =& \frac{1}{2\sqrt{3}} \left(ZQ_{\mathrm{central}}^\mathrm{A}  + DQ_{\mathrm{central}}^\mathrm{A} + \frac{1}{\sqrt{2}} \left(ZQ_{\mathrm{central}}^\mathrm{\Sigma E} + DQ_{\mathrm{central}}^\mathrm{\Sigma E} \right) \right)
    \end{aligned}
   \label{eq:TransitionTugarinov}
\end{equation}

\subsection{Relaxation mechanisms}
Nuclear spin relaxation is described using the Bloch-Redfield-Wangsness (BRW) relaxation theory, comprehensive descriptions of which can be found elsewhere.\cite{kumar_cross-correlations_2000,goldman_interference_1984, kowalewski_nuclear_2006, nicholas_nuclear_2010,abragam_principles_1961} Calculations were performed using the framework of SpinDynamica.\cite{Bengs_MagnResonChecm_2017} We considered the three \Proton-\Proton~ and three \Carbon-\Proton~DD interactions, as well as the \Carbon-CSA interaction, when mentioned.

\subsection{Spectral density functions}
\begin{table*}
	\caption{Spectral density functions used in the description of relaxation in a methyl group and associated values of $S_m^2(\theta_{\vec{\imath},\vec{\jmath}})$ and $\mathcal{P}_2(\cos(\theta_{\vec{\imath},\vec{\jmath}})$ for different interactions. $\JCCH$ and $\JCCHH$ are negative at all magnetic fields. CSA: Chemical Shift Anisotropy. DD: dipole-dipole.}
	\begin{center}
		{\def\arraystretch{1.3}
		\begin{tabular}{ccccc}
			Notation & Correlation & Interactions & $S_m^2(\theta_{\vec{\imath},\vec{\jmath}})$  & $\mathcal{P}_2(\cos(\theta_{\vec{\imath},\vec{\jmath}})$ \\
			\hline
		 	$\JCC$ & auto-correlation C-C  & CSA & 1  & 1   \\%
			$\JCH$ & auto-correlation C-H  & DD & 1/9  & 1   \\%
			$\JHH$ & auto-correlation H-H  & DD & 1/4  & 1   \\
		 	$\JHCH$ & cross-correlation between two C-H pairs &  DD/DD & 1/9  &  -1/3   \\%
		 	$\JHHH$ & cross-correlation between two H-H pairs & DD/DD &  1/4 & -1/8     \\%
			$\JCHH$ & cross-correlation between a C-$\mathrm{H}_i$ and a $\mathrm{H}_i$-$\mathrm{H}_j$ pair &  DD/DD & 1/6 & 1/2     \\%
			$\JCHHH$ & cross-correlation between a C-$\mathrm{H}_i$ and a $\mathrm{H}_j$-$\mathrm{H}_k$ pair &  DD/DD & 1/6 & -1/2     \\
			$\JCCH$ & cross-correlation between the \Carbon-CSA and a C-H pair &  CSA/DD & -1/3 & -1/3     \\
			$\JCCHH$ & cross-correlation between the \Carbon-CSA and a H-H pair &  CSA/DD & -1/2 & -1/2     \\
		\end{tabular}
		}
	\end{center}
	\label{Table:ParamVal}
\end{table*}
To take into account the methyl rotation around its symmetry axis, the model-free approach \cite{lipari_model-free_1982} has been modified to include three types of motions, considered to be uncorrelated: the global tumbling, motions of the methyl symmetry axis occuring on the nano- to sub-nano second time scales,\cite{Clore_JACS_1990} and the rotation of the methyl group, resulting in a correlation function similar to a form previously introduced for methyl groups:\cite{Liao_JPC_2012}
\begin{equation}
	\begin{aligned}
    	C(t, \theta_{\vec{\imath},\vec{\jmath}}) =& \frac{1}{5}e^{-t/\TAUC}(S_f^2 + (1-S_f^2)e^{-t/\TAUF})\times (S_m^2 (\theta_{\vec{\imath},\vec{\jmath}}) + (\mathcal{P}_2(cos(\theta_{\vec{\imath},\vec{\jmath}}))-S_m^2 (\theta_{\vec{\imath},\vec{\jmath}}))e^{-t/\TAUM})
    \end{aligned}
\end{equation}
where $\TAUC$ is the overall global tumbling correlation time, $S_f^2$ and $\TAUF$ are the order parameter and correlation time for the motions of the symmetry axis aligned with the CC bond, and $\TAUM$ is the correlation time for the rotation of the methyl group. $S_m^2(\theta_{\vec{\imath},\vec{\jmath}})$ is the order parameter of the methyl group which can be expressed as $S^2_m (\theta_{\vec{\imath},\vec{\jmath}}) = \mathcal{P}_2[\cos(\theta_{\vec{\imath},\vec{CC}})] \times \mathcal{P}_2[\cos(\theta_{\vec{\jmath},\vec{CC}})]$,\cite{Frueh_PNMR_2002} where $\mathcal{P}_2$ is the second order Legendre polynomial and $\vec{CC}$ is the vector aligned along the C-C bond and associated with the symmetry axis of the system. $\theta_{\vec{\imath},\vec{\jmath}}$ defines the angle between the vectors \textit{i} and \textit{j} formed by the two pairs of nuclei involved in the considered DD interactions, or the symmetry axis of the CSA tensor, and allows for possible cross-correlation. The Fourier transform of the correlation function gives the following spectral density function for the Model-Free for Methyl (MFM) model:
\begin{equation}
	\begin{aligned}
		\mathcal{J}_{MFM}(\omega, \theta_{\vec{\imath},\vec{\jmath}}) =& 
        \cfrac{2}{5}  [ S_m^2 (\theta_{\vec{\imath},\vec{\jmath}}) (S_f^2L(\omega, \TAUC)  + (1-S_f^2)L(\omega, \TAUF'))  \\
        & +(\mathcal{P}_2(\cos(\theta_{\vec{\imath},\vec{\jmath}}))-S_m^2 (\theta_{\vec{\imath},\vec{\jmath}})) \times (S_f^2 L(\omega, \TAUM') +  (1-S_f^2) L(\omega ,\TAUF''))]
    \end{aligned}
    \label{eq:SpectralDensityJMFM}
\end{equation}
where $\omega$ is the Larmor frequency, effective correlation times are expressed as $\tau_k'^{-1}=\tau_k^{-1}+\TAUC^{-1}$, ($k \in \{f, \mathrm{met}\}$) and $\tau_k''^{-1}=\tau_k^{-1}+\TAUC^{-1}+\TAUM^{-1}$, ($k \in \{f\}$), and $L(\omega, \tau)=\tau/(1+(\omega \tau)^2)$ stands for the Lorentzian function. This spectral density function will be used throughout the text, unless otherwise specified. For the sake of simplicity a compact notation is used in the rest of the manuscript. Spectral density functions are labeled with indices referring to the auto- and cross-correlated interactions following a notation suggested by Werbelow and Grant.\cite{Werbelow_JCP_1975} Notations used for the spectral density functions and values of $S_m^2(\theta_{\vec{\imath},\vec{\jmath}})$ and $\mathcal{P}_2(\theta_{\vec{\imath},\vec{\jmath}})$ are listed in Table\,\ref{Table:ParamVal}.\\
In the hypothesis of an infinitely fast methyl rotation (IFR) the last two terms in Eq.\,\ref{eq:SpectralDensityJMFM} vanish, leading to the spectral density function $\mathcal{J}_{MFM}^{IFR}$:
\begin{equation} 
	\begin{aligned}
		\mathcal{J}_{MFM}^{IFR}(\omega, \theta_{\vec{\imath},\vec{\jmath}}) =&  \lim_{\TAUM \rightarrow 0}  \mathcal{J}_{MFM}(\omega, \theta_{\vec{\imath},\vec{\jmath}}) \\
        =& \cfrac{2}{5}  \left[ S_m^2 (\theta_{\vec{\imath},\vec{\jmath}}) S_f^2 L(\omega, \TAUC) +  S_m^2 (\theta_{\vec{\imath},\vec{\jmath}}) (1-S_f^2) L(\omega, \TAUF') \right]
    \end{aligned}
    \label{eq:SpectralDensityIF}
\end{equation}
Importantly, under the infinitely fast methyl rotation approximation, the dependence on the relative orientation of all interactions vanishes, leading to $\JHCH = \JCH$ and $\JHHH = \JHH$ which is important for relaxation interference (see below). Finally, the \SlowTumbling~approximation implies $\mathcal{J}(\omega)=0$ for $\omega \ne 0$.  \\
Tugarinov \emph{et al.} used a simpler form of spectral density function adapted to slow tumbling for the overall rotational diffusion and Infinitely Fast Internal Motions (IFIM), in which all internal correlation times are zero:\cite{tugarinov_cross-correlated_2003}
\begin{equation} 
	\begin{aligned}
		\mathcal{J}_{MFM}^{IFIM}(\omega, \theta_{\vec{\imath},\vec{\jmath}}) =& \cfrac{2}{5}  S_m^2 (\theta_{\vec{\imath},\vec{\jmath}}) S_f^2 L(\omega, \TAUC)
    \end{aligned}
    \label{eq:SpectralDensityIFIM}
\end{equation}
In the analysis presented here, the following parameters $S_f^2=0.5$, $\TAUC$=10\,ns, $\TAUF$=100\,ps and $\TAUM$=5\,ps will be used, if not specified otherwise.\cite{Cousin_JACS_2018} Introducing the additional correlation time $\TAUF$ for internal motions does not change the general features of methyl TROSY. Supplementary Material contains results for calculations performed using a correlation time typical for a large macromolecule: $\TAUC$=100\,ns.

%% file: Section/4_HighField.tex
\section{Methyl-TROSY at high field}
\begin{figure}
	\begin{center}
		\includegraphics[width=0.7\textwidth]{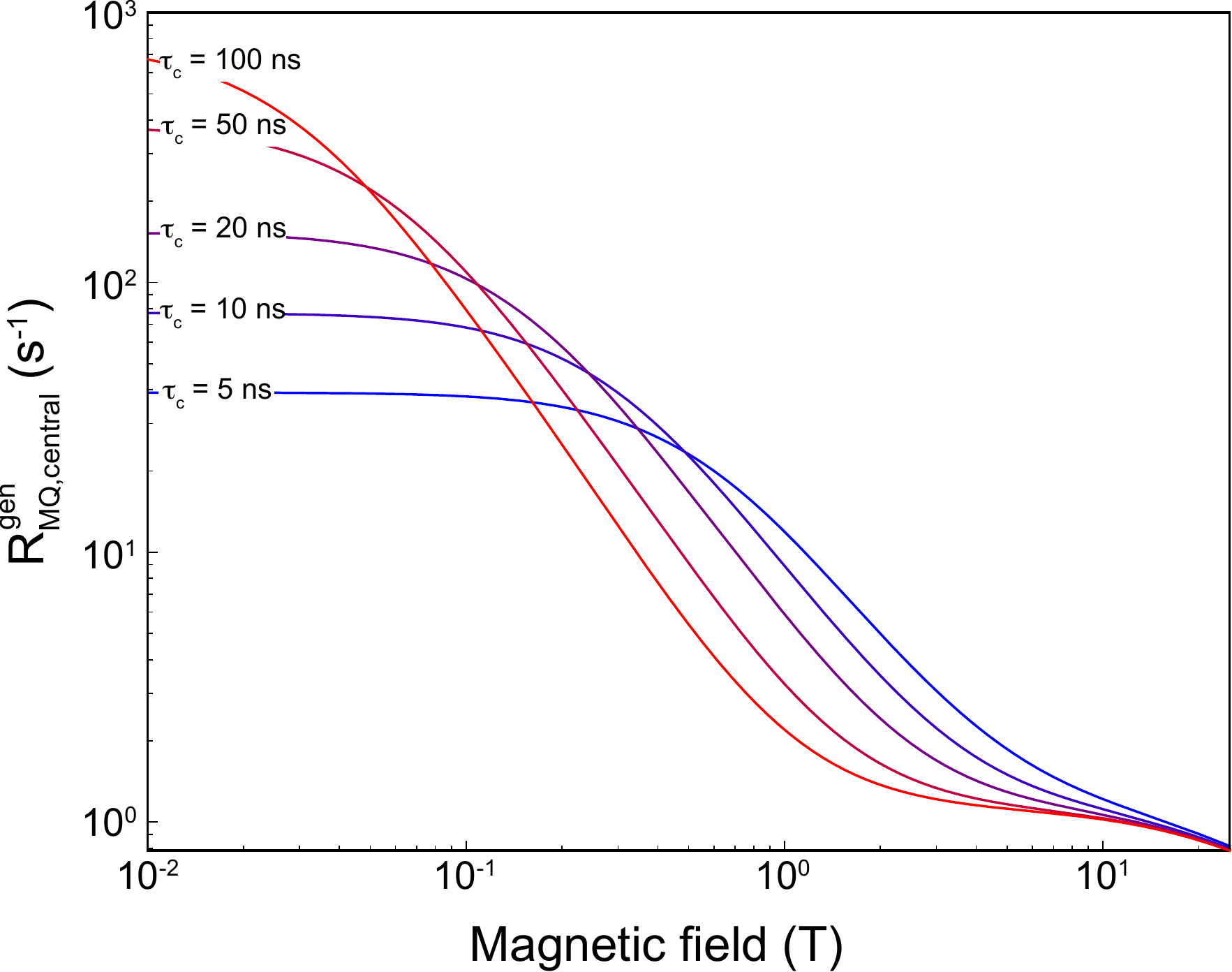}
	\end{center}
	\caption{Variation of the general expression of the central line relaxation rate $R_{\mathrm{MQ,central}}^{gen}$ with respect to the magnetic field for different values of the overall corelation time $\TAUC$. Relaxation rates were calculated using the spectral density function $\mathcal{J}_{MFM}$ and Eq.\,\ref{eq:RNoAssump}. The contribution of the CSA to the relaxation is not included.}
	\label{fig:R2in}
\end{figure}

In the absence of the proton refocusing pulse during the indirect evolution period of the HMQC pulse sequence, the coupling between the evolving \Carbon-\Proton~spin pair and the passive \Proton-\Proton~spin pair leads to a triplet arising from the different spin states of the passive protons (Fig.\,\ref{fig:BasisMethyl}.b). The central line of the triplet is much sharper than the outer lines (for a detailed description of the HMQC spectrum, see \cite{tugarinov_cross-correlated_2003}). The dipolar contributions to the transverse relaxation rates for the outer and central single-transition operators ($T_{outer}$ and $T_{central}$ defined in Eq.\,\ref{eq:TransitionTugarinov}) can be expressed as general (gen) expressions $R_{\mathrm{MQ,outer}}^{\mathrm{gen}}$ and $R_{\mathrm{MQ,central}}^{\mathrm{gen}}$:
\begin{equation}
	\begin{aligned}
		R_{\mathrm{MQ,outer}}^{\mathrm{gen}} =&  \frac{1}{8} d_{\sCH}^2 \left[ 8 \JCH(0) + 9 \JCH(\OMc) +3\JCH(\OMc-\OMh) + 9 \JCH(\OMh) + 18 \JCH(\OMc+\OMh)  \right] \\
	& + \frac{1}{4} d_{\sCH}^2 \left[ 4 \JHCH(0) + 3 \JHCH(\OMc) + \JHCH(\OMc-\OMh) + 3 \JHCH (\OMh) + 6 \JHCH(\OMc+\OMh) \right] \\
	& + \frac{3}{4} d_{\sHH}^2 \left[ 3 \JHH(0) + 4 \JHH(\OMh) + 2 \JHH(2 \OMh) \right] \\
	& + \frac{3}{4} d_{\sHH}^2 \left[ 3 \JHHH(0) + 2 \JHHH(\OMh) - 2 \JHHH(2 \OMh) \right] \\
		R_{\mathrm{MQ,central}}^{\mathrm{gen}} =&  \frac{1}{8} d_{\sCH}^2 \left[ 8 \JCH(0) + 9 \JCH(\OMc) +  3 \JCH(\OMc-\OMh) + 9 \JCH(\OMh) + 18 \JCH(\OMc+\OMh) \right] \\
	& + \frac{1}{4} d_{\sCH}^2 \left[ -4 \JHCH(0) - 3 \JHCH(\OMc) +  \JHCH(\OMc-\OMh) + 3 \JHCH(\OMh) + 6 \JHCH(\OMc+\OMh) \right] \\
	& + \frac{3}{4} d_{\sHH}^2 \left[ 3 \JHH(0) + 4 \JHH(\OMh) + 2 \JHH(2 \OMh) \right] \\
	& + \frac{3}{4} d_{\sHH}^2 \left[ -3 \JHHH(0) + 2 \JHHH(2 \OMh) \right]
     \label{eq:RNoAssump}
     \end{aligned}
\end{equation}
where $d_{ij}$ are dipolar coefficients for the DD interaction between nuclei $i$ and $j$, $d_{ij}=-\frac{\mu_0 \hbar}{4\pi}\frac{\gamma_i \gamma_j}{r_{ij}^3}$ with $\mu_0$ the permeability of free space, $\hbar$ the Planck's constant divided by $2\pi$, $\gamma_n$ the gyromagnetic ratio of nucleus \textit{n}, and $r_{ij}$ the distance between nuclei \textit{i} and \textit{j}. Importantly, the rates $R_{\mathrm{MQ,central}}^{\mathrm{gen}}$ and $R_{\mathrm{MQ,outer}}^{\mathrm{gen}}$ are not equal due to different cross-correlated contributions depending on $\JHCH$ and $\JHHH$. Using the \SlowTumbling~approximation, we only retain terms of the spectral density function evaluated at 0 frequency, leading to:
\begin{equation}
	\begin{aligned}
		R_{\mathrm{MQ,outer}}^{ST} =& d_{\sCH}^2 \left[ \JCH(0) + \JHCH(0)  \right]  + \frac{9}{4} d_{\sHH}^2 \left[ \JHH(0) + \JHHH(0) \right] \\
		R_{\mathrm{MQ,central}}^{ST} =& d_{\sCH}^2 \left[ \JCH(0) - \JHCH(0) \right]  +  \frac{9}{4} d_{\sHH}^2 \left[ \JHH(0) -  \JHHH(0) \right]
    \label{eq:ExtremeNarrowing}
    \end{aligned}
\end{equation}
As explained in the previous section, under infinitely fast methyl rotation, the spectral density functions for auto- and cross-correlation are equal, $\JCH=\JHCH$ and $\JHH = \JHHH$, so that:
\begin{equation} 
	\begin{aligned}
		R_{\mathrm{MQ,outer}}^{ST, IFR} &= 2  d_{\sCH}^2 \JCH(0) + \frac{9}{2} d_{\sHH}^2\JHH(0) \\
		R_{\mathrm{MQ,central}}^{ST, IFR} &= 0
        \label{eq:RSimplification}
        \end{aligned}
\end{equation}
The complete cancellation of the relaxation rate of the central line arises from the combination of two approximations: \SlowTumbling~and infinitely fast rotation of the methyl group. A similar cancellation of auto- and cross-correlated relaxation terms is responsible for the existence of long-lived nuclear spin states in methyl groups.\cite{Dumez_JCP_2015} Introducing the spectral density function $\mathcal{J}_{MFM}^{IFIM}$ (Eq.\,\ref{eq:SpectralDensityIFIM}), we obtain:
\begin{equation}
	\begin{aligned}
		R_{\mathrm{MQ,outer}}^{ST, IFIM} &= \frac{\mu_0}{4\pi} \left(2 \frac{2}{5}\frac{S_m^2(\theta_{\vec{CH}, \vec{CH}}) S_{f}^2 \gammaH^2 \gammaC^2 \hbar^2 \TAUC}{r_{\mathrm{HC}}^6} + \frac{9}{2} \frac{2}{5} \frac{S_m^2(\theta_{\vec{HH}, \vec{HH}}) S_{f}^2 \gammaH^4\hbar^2 \TAUC}{r_{\mathrm{HH}}^6} \right) \\%
		R_{\mathrm{MQ,central}}^{ST, IFIM}&=0
    \label{eq:RInTuga}
   	\end{aligned}
\end{equation}
Replacing $S_m^2$ according to Table\,\ref{Table:ParamVal}, we obtain the same expressions as reported in \cite{tugarinov_cross-correlated_2003}:
\begin{equation}
	\begin{aligned}
		R_{\mathrm{MQ,outer}}^{ST, IFIM} &= \frac{\mu_0}{4\pi} \left(\frac{4}{45}\frac{S_{f}^2 \gammaH^2 \gammaC^2 \hbar^2 \TAUC}{r_{\mathrm{HC}}^6} + %
	\frac{9}{20} \frac{S_{f}^2 \gammaH^4\hbar^2 \TAUC}{r_{\mathrm{HH}}^6} \right) \\%
		R_{\mathrm{MQ,central}}^{ST, IFIM}&=0
    \label{eq:RInTuga}
   	\end{aligned}
\end{equation}
The relaxation rate of the outer lines calculated using the general expression $R_{\mathrm{MQ,outer}}^{\mathrm{gen}}$ (Eq.\,\ref{eq:RNoAssump}) and in the \SlowTumbling~approximation $R_{\mathrm{MQ,outer}}^{ST}$ (Eq.\,\ref{eq:ExtremeNarrowing}) calculated using the spectral density function $\mathcal{J}_{MFM}$ compare well for magnetic fields higher than 5\,T, both of them being independent of the magnetic field and proportional to the global tumbling correlation time $\TAUC$ (Supplementary Material Fig.\,S1.a and Fig.\,S2.a). The value of the relaxation rate of the central line $R_{\mathrm{MQ,central}}^{ST}$ approaches zero in the \SlowTumbling~approximation (Eq.\,\ref{eq:ExtremeNarrowing} and Supplementary Material Fig.\,S1.b), as analytically calculated in the case of the infinitely fast rotation of the methyl group (Eq.\,\ref{eq:RSimplification}). In the general case, the central line relaxation rate $R_{\mathrm{MQ,central}}^{\mathrm{gen}}$ has a non-zero value. For magnetic fields higher than 5\,T, it is small and can be considered independent of the magnetic field and global tumbling correlation time $\TAUC$ (Supplementary Material Fig.\,S2.b), thus reproducing the expected behavior predicted by the \SlowTumbling~approximation. These calculations show that in the frame of the operator expansion introduced by Tugarinov \emph{et al.} \cite{tugarinov_cross-correlated_2003} (Eq.\,\ref{eq:TransitionTugarinov}), the \SlowTumbling~approximation allows an accurate description of the relaxation rates of the triplet at moderate and high magnetic fields.\\
However, using this operator expansion, calculation of the relaxation rate of the central line of the triplet shows a rapid increase of $R_{\mathrm{MQ,central}}^\mathrm{gen}$ at low magnetic fields (Fig.\,\ref{fig:R2in}). Such an increase is in contradiction with the favorable relaxation properties of multiple-quantum coherences at 0.33\,T in the 2F-HZQC experiment recorded on a sample of \ubiquitin.\cite{cousin_recovering_2016} As $R_{\mathrm{MQ,central}}^\mathrm{gen}$ is calculated without making the initial hypotheses of \SlowTumbling~and infinitely fast rotation of the methyl group, this discrepancy between theory and experiment cannot be attributed to the expression of the spectral density function but to the expansion of operators employed. In the following section, we will show that the expansion of operators introduced in the theoretical framework section allows us to understand the relaxation properties of MQ coherences at low magnetic fields.

%% file: Section/5_Generalization.tex
\section{Methyl-TROSY beyond the slow tumbling limit}
\subsection{Definition of a suitable basis}
So far, calculations were done using an expansion of the single-transition operators between those contributing to the sharp central line, and those contributing to the broad outer lines of the triplet. Here, we suggest to calculate independentely the relaxation properties of each single-transition operator in order to identify slowly relaxing components. Suitable bases for the expansion of the \CpHm~and \CpHp~coherences are presented in the theoretical framework section (Table\,\ref{Table:LabelOp} and Eq.\,\ref{eq:MQdecomp}).

\subsection{Identification of a slowly relaxing term in the new operator basis}
\begin{figure}
	\begin{center}
		\includegraphics[width=0.9\textwidth]{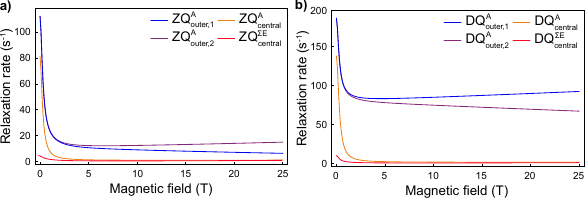}
	\end{center}
	\caption{Magnetic field variation of the relaxation rates of the four single-transition operators of the MQ triplet in the zero-quantum basis (\textbf{a}) $\mathcal{B}_{ZQ}$ and the double-quantum basis (\textbf{b}) $\mathcal{B}_{DQ}$. Relaxation rates were calculated using the spectral density function $\mathcal{J}_{MFM}$.}
	\label{fig:CalculationRateC3basis}
\end{figure}
The contribution of the carbon-13 Chemical Shift Anisotropy (CSA) is included in the following analysis. It is assumed to be axially symmetric with anisotropy $\Delta \sigma = \sigma_\mathrm{zz} - \left(\sigma_\mathrm{xx} + \sigma_\mathrm{yy}\right)/2= 20\,\mathrm{ppm}$ and aligned with the direction of the C-C bond. The proton CSA is expected to be approximately 1\,ppm \cite{Tugarinov_JBNMR_2004} and is neglected in our analysis. Relaxation rates for the zero- and double-quantum single-transition operators forming the bases $\mathcal{B}_{ZQ}$ and $\mathcal{B}_{DQ}$ are shown in Fig.\,\ref{fig:CalculationRateC3basis}. The single-transition operators contributing to the outer lines of the triplet relax faster than the single-transition operators of the central line, at all magnetic fields (between 0.01\,T and 25\,T), in agreement with the previous analysis.\cite{tugarinov_cross-correlated_2003} Interestingly, the two operators corresponding to the outer lines do not have the same relaxation properties at high fields. This effect arises from the CSA/DD contributions to relaxation, which are negligible at low field (see Supplementary Material for expression of the relaxation rates).\\
The two operators that describe the central line have comparable relaxation properties at high fields (higher than ca. 5\,T). This is consistent with the treatment proposed by Tugarinov \emph{et al.} in the slow tumbling regime with infinitely fast methyl rotation.\cite{tugarinov_cross-correlated_2003} At lower fields, the two single-transition operators of the central line have drastically different relaxation behaviors.This difference arises even in the infinitely fast methyl-group rotation limit (Supplementary Material Fig.\,S3). The A-manifold operators contributing to the central line $ZQ_{\mathrm{central}}^\mathrm{A}$ and $DQ_{\mathrm{central}}^\mathrm{A}$ relax much faster while the sums of the E single-transition operators still relax slowly. These predictions could be experimentally verified by separating the \Proton~transitions of the two manifolds using a two-field version of the approach described in \cite{Tugarinov_JACS_2007_bis}.\\
The ratio of the relaxation rates of the two operators corresponding to the central line is around 10 in the ZQ and DQ cases at 0.33\,T with the parameters for dynamics used here (see theoretical framework section). Similar conclusions can be drawn in the case of large proteins ($\TAUC$=100\,ns) where the \SlowTumbling~approximation is almost justified at 0.33\,T. The sums of the E single-transition operators have small relaxation rates around 4\,s$^{-1}$ in $\mathcal{B}_{ZQ}$ and 5\,s$^{-1}$ in $\mathcal{B}_{DQ}$ at 0.33\,T (Supplementary Material Fig.\,S4.a and b.).The A-manifold central MQ transition operators $ZQ_{\mathrm{central}}^\mathrm{A}$ and $DQ_{\mathrm{central}}^\mathrm{A}$ have relaxation rates higher but comparable to their E-manifold equivalents at 0.33\,T (11\,s$^{-1}$ in $\mathcal{B}_{ZQ}$ and 20\,s$^{-1}$ in $\mathcal{B}_{DQ}$). This shows that the combined operators approach \cite{tugarinov_cross-correlated_2003} is still valid at relatively low magnetic fields for high molecular-weight proteins. \\
Taken together, these results suggest that the Methyl-TROSY effect is retained at low fields for only one of the two single-transition operators that contribute to the central line of the triplet, while both single-transition operators relax slowly at high field.

\subsection{Cross relaxation between and within the lines}
\begin{figure}
	\begin{center}
		\includegraphics[width=0.9\textwidth]{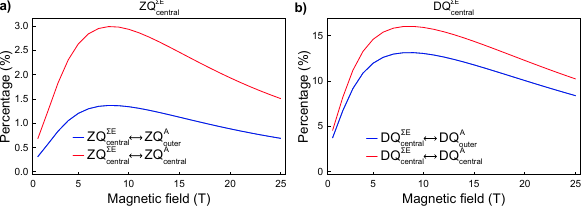}
	\end{center}
	\caption{Magnetic field variation of the percentage of cross-relaxation between the E- and A-manifolds of the MQ triplet in the zero-quantum basis (\textbf{a}) $\mathcal{B}_{ZQ}$ and the double-quantum basis (\textbf{b}) $\mathcal{B}_{DQ}$. The cross-relaxation rates are expressed as the percentage of the auto-relaxation rate of the (\textbf{a}) $ZQ_{\mathrm{central}}^\mathrm{\Sigma E}$ and (\textbf{b}) $DQ_{\mathrm{central}}^\mathrm{\Sigma E}$ transition operators. Relaxation rates were calculated using the spectral density function $\mathcal{J}_{MFM}$. The two single-transition operators contributing to the outer lines of the triplet have the same cross-relaxation rates with the E-manifold in both the ZQ and DQ cases.}
	\label{fig:CrossRelaxation}
\end{figure}
In order to describe the relaxation of methyl groups comprehensively, we studied the cross-relaxation pathways related to the slow-relaxing terms. This requires a full treatment of the relaxation matrix in the considered basis, and is essential to ensure that the single-transition operators considered above are good approximations of the eigenvectors of the relaxation matrix. Two types of cross-relaxation pathways have to be considered: cross-relaxation between the slowly relaxing single-transition operator of the central line and the fast relaxing single-transition operator of the central line (intra-line transfer) or cross-relaxation between the central line of the triplet and the outer lines (inter-line transfer). Percentage of intra- and inter-line cross-relaxation with respect to the auto-relaxation of the $MQ_{\mathrm{central}}^\mathrm{\Sigma E}$ operators are shown in Fig.\,\ref{fig:CrossRelaxation}.\\
Inter-line cross-relaxation rates appear as non-secular in the interaction frame of \Carbon-\Proton~scalar coupling interactions (they oscillate at the angular frequency $2\pi$J with J the scalar coupling constant). Thus the predicted cross-relaxation rates with the single-transition operators contributing to the outer lines of the triplet (blue curves in Fig.~\ref{fig:CrossRelaxation}) have no effect on the relaxation of the term contributing to the central line. On the other hand, intra-line cross-relaxation is always secular. At high field (higher than 5T) cross-relaxation has no effect as the A- and E-manifold central coherences have the same relaxation rates (Fig.\,\ref{fig:CalculationRateC3basis}). At low fields (lower than 1 T), the $ZQ_{\mathrm{central}}^\mathrm{A}$ and $DQ_{\mathrm{central}}^\mathrm{A}$ terms relax much faster than the sums of the E single-transition operators. The $ZQ_{\mathrm{central}}^\mathrm{\Sigma E}$ cross-relaxation rate with $ZQ_{\mathrm{central}}^\mathrm{A}$ ranges from 0.5 to 3\% of the auto-relaxation of the slowly relaxing $ZQ_{\mathrm{central}}^\mathrm{\Sigma E}$ from 1 to 5\,T so that cross-relaxation effects can be neglected (red curve in Fig.\,\ref{fig:CrossRelaxation}.a). The intra-line cross-relaxation rate represents 5 to 15\% of the auto-relaxation rate in the DQ case (Fig.\,\ref{fig:CrossRelaxation}.b), and may have a small effect on the decay of the polarization. Interestingly, intra-line cross-relaxation rates are independent of the magnetic field as they only depend on spectral density function evaluated at zero-frequency (see Supplementary Material for detailed relaxation rate expressions). Moreover, the percentage of cross-relaxation is independent of the global tumbling correlation time $\TAUC$ and our conclusions can be extended to large proteins.

\subsection{Fast methyl rotation is important for a TROSY effect}
\begin{figure*}
	\begin{center}
		\includegraphics[width=0.8\textwidth]{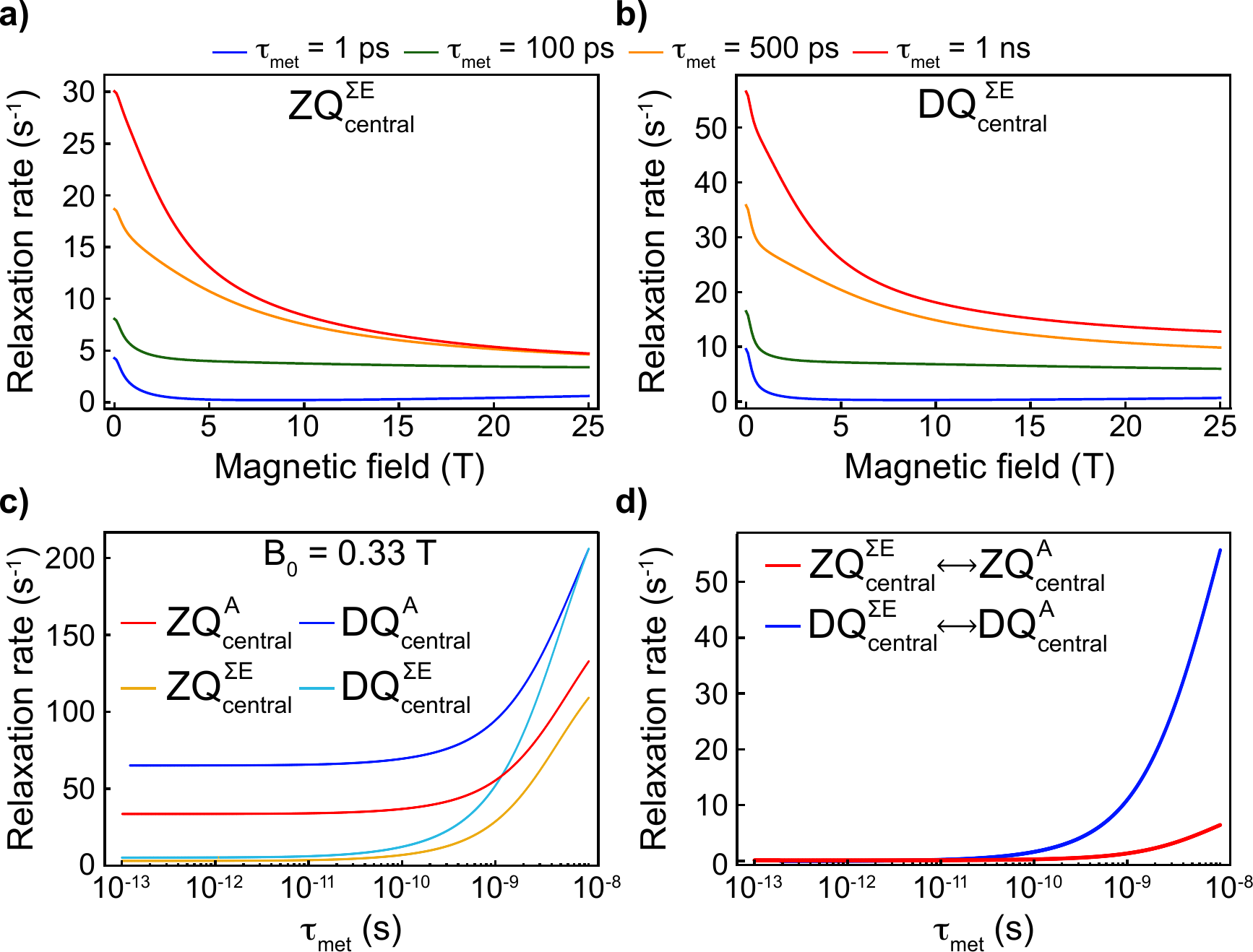}
	\end{center}
	\caption{Effect of methyl group rotation on methyl relaxation properties. Evolution of the relaxation rates for the operators $ZQ_{\mathrm{central}}^\mathrm{\Sigma E}$ (\textbf{a}) and $DQ_{\mathrm{central}}^\mathrm{\Sigma E}$ (\textbf{b}) from 0.1 to 25\,T with $\TAUM$ values ranging from 1\,ps to 1\,ns. \textbf{c.} Evolution of the auto-relaxation rates of the operators contributing to the central line of the triplet as a function of the methyl group rotation correlation time $\TAUM$ at 0.33\,T. \textbf{d.} Evolution of the intra-line cross-relaxation rates as a function of the methyl group rotation correlation time $\TAUM$. These rates are independent of the magnetic field as can be seen in the expressions in the Supplementary Material. Relaxation rates were calculated using the spectral density function $\mathcal{J}_{MFM}$.}
	\label{fig:TmEffect}
\end{figure*}
As discussed in the previous section, the theory of methyl TROSY proposed by Tugarinov \emph{et al.} is based on the hypotheses of slow tumbling of the protein and infinitely fast rotation of the methyl group. We investigated the effect of the finite speed of rotation of the methyl group on both auto- and cross-relaxation rates.\\
Fast rotation of the methyl group is essential for favorable auto-relaxation rates as even the previously identified slowly relaxing terms have significantly higher relaxation rates for rotation correlation times $\TAUM$ values larger than 1\,ns, especially at low fields (Fig.\,\ref{fig:TmEffect}.a and b). At 0.33\,T, the auto-relaxation rates of the operators for the central line of the triplet are mostly independendent of the correlation time $\TAUM$ in the range from 0.1\,ps (close to the infinitely fast methyl rotation) to 100\,ps, \textit{i.e.} the chosen value for the correlation time $\TAUF$ (Fig.\,\ref{fig:TmEffect}.c). This correlates with a higher loss of polarization through intra-line cross-relaxation as the slow and fast relaxing terms are mixed rapidely (Fig.\,\ref{fig:TmEffect}.d). In agreement with the initial treatment of Tugarinov \emph{et al.}, a fast rotating group on the pico-second to few tens of pico-second time scales ensures an efficient methyl-TROSY effect at high fields as well as the ability to record a methyl-TROSY spectrum at low magnetic fields (below 1\,T). The same conclusions can be drawn for larger proteins (Supplementary Material Fig.\,S4.c-f).\\
In the case of protein NMR, isoleucine is a favorable methyl group-bearing residue with low energy barriers for methyl group rotation.\cite{Cousin_JACS_2018} The rotation of the methyl group in alanine can be significantely affected by interactions with the protein backbone, leading to a higher $\TAUM$.\cite{Batchelder_JACS_1983} Hence, signals from \CHH groups in alanine side-chains are expected to be broader than signals of isoleucines, or even leucines or valines.

\subsection{What happens at very high fields}
\begin{figure}
	\begin{center}
		\includegraphics[width=0.9\textwidth]{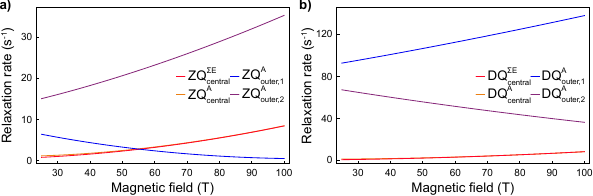}
	\end{center}
	\caption{Effect of the \Carbon-CSA contribution on the auto-relaxation rates. Depence of the auto-relaxation rates of the single-transition operators contributing to the central and outer lines of the triplet on the magnetic field for very high fields (from 25 to 100\,T) in the basis $\mathcal{B}_{ZQ}$ (\textbf{a}) and $\mathcal{B}_{DQ}$ (\textbf{b}). Relaxation rates were calculated using the spectral density function $\mathcal{J}_{MFM}$.}
	\label{fig:CSAEffect}
\end{figure}
The contribution of the chemical shift anisotropy to relaxation is negligible at the low fields considered so far. CSA contributions to the relaxation of MQ transition operators increase with the magnetic field. At the highest currently commercially available magnetic field (B$_0$ = 23.5\,T), this contribution to relaxation is not negligible but still small. We have investigated how relaxation due to the CSA interaction would alter HMQC spectra of methyl groups at magnetic fields that will be commercially available in the future (B$_0$ $>$ 25\,T). As discussed above, the proton CSA is neglected.\\
The contribution of the chemical shift anisotropy leads to a small but significant increase of the auto-relaxation rates for the single-transition operators of the central line (Fig.\,\ref{fig:CSAEffect} and Supplementary Material Fig.\,S4.g-h). Such increase is expected moderately deteriorate the quality of HMQC spectra. \\
By contrast, the relaxation rates of the outer lines of the triplet change dramatically with the magnetic field in both the ZQ and DQ cases. When the magnetic field increases, if the passive spins (\Proton) are in the $\alpha$ state (operators $\mathrm{MQ}^A_{\mathrm{outer,1}}$), auto-relaxation rates decrease for ZQ transition operators but increase for DQ transition operators. On the other hand, when passive protons spins are in the $\beta$ state (operators $\mathrm{MQ}^A_{\mathrm{outer,2}}$), the auto-relaxation rates increase for ZQ transition operators but decrease DQ transition operators. The magnetic-field variation of relaxation rates is dominated by the interference between the \Carbon-CSA and the \Proton-\Proton~DD interactions. The additional \Carbon-CSA/\Carbon-\Proton~DD cross-correlated contribution leads to a stronger field-dependence of relaxation rates for DQ transition operators than for ZQ transition operators as can be infered from analytical expressions (Supplementary Material, remembering that $\JCCH$ and $\JCCHH$ are negative at all magnetic fields). The overall effect of CSA/DD cross-correlated relaxation leads to a crossing of the field-dependence of relaxation rates for ZQ transition operators at very high fields: the single-transition operators $\mathrm{ZQ}^A_{\mathrm{outer,1}}$ (passive spins in $\alpha$ state) becomes the single-transition operator with the smallest auto-relaxation rate for magnetic fields higher than ca. 55\,T (2.3\,GHz). For the central line, the sum of the CSA/DD cross-correlations vanishes so that the field variation of CSA constributions to relaxation is only due to the auto-correlation (see Supplementary Material for expressions of relaxation rates). For large proteins at very high fields, recording a HZQC spectrum \cite{tugarinov_line_2004} on the $\mathrm{ZQ}^A_{\mathrm{outer,1}}$ appears to be more favorable, introducing a new form of methyl-TROSY based on cancelation between CSA and DD contributions to relaxation. This prediction suggests that transverse relaxation-optimized spectroscopy due to CSA/DD interference should be investigated in aliphatic groups at fields $B_0>$25\,T.

%% file: Section/6_ComparisonExp.tex
\section{Two-field HZQC analysis}
\begin{figure}
	\begin{center}
		\includegraphics[width=0.9\textwidth]{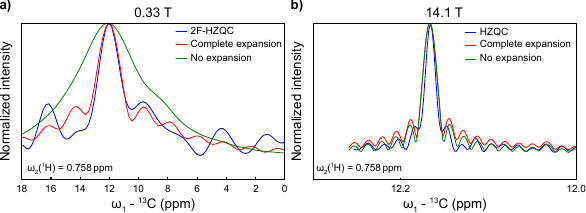}
	\end{center}
	\caption{Simulation of a cross-section along the indirect dimension of the 2F-HZQC spectrum of Ile-30 from \ubiquitin~measured with $t_1$ evolution at (\textbf{a}) 0.33\,T or (\textbf{b}) a single-field HZQC recorded at 14.1\,T. The cross-sections from the experimental spectra (blue) are compared with a simulation using the complete expansion of the basis (red) and using the previously reported expression of relaxation rate $R_{\mathrm{MQ,central}}^{\mathrm{gen}}$ (green). Each cross-section of each spectrum is normalized independently so that maximum intensities are identical.}
	\label{fig:HZQCsimu}
\end{figure}
We have recently introduced two-field NMR spectroscopy, where different parts of a pulse sequence take place at different magnetic fields.\cite{cousin_recovering_2016, cousin_high-resolution_2016, Kaderavek_JCPC_2017} These experiments were carried out on a two-field NMR spectrometer that consists of two magnetic centers operating at 14.1\,T and 0.33\,T, coupled by a sample shuttle.\cite{Charlier_JACS_2013, cousin_high-resolution_2016}  In particular, we have recorded two-dimensional spectra of small molecules with indirect evolution at 0.33\,T and direct proton detection at 14.1\,T.\cite{cousin_high-resolution_2016}  High-resolution spectra were obtained, in spite of residual field inhomogeneities, by exploiting the favorable properties of zero-quantum coherences which are insensitive to magnetic field inhomogeneity. In the case of heteronuclear spin systems, additional tailoring of effective Hamiltonians was necessary to obtain full compensation of B$_0$ inhomogeneities. In this two-field heteronuclear zero-quantum coherence experiment (2F-HZQC) the density operator that evolves at 0.33\,T is partly \Proton-\Carbon~zero-quantum coherence, partly \Proton-\Carbon~double-quantum coherence. In order to ensure the full cancellation of magnetic field inhomogeneity, the evolution is tailored by $\pi$-pulses to obtain an effective propagator equal to: 
\begin{equation}
	\hat{U}_{LF}(t)=\exp \left[ in_1 t_1 \left( \frac{\gamma_{\mathrm{C}}}{\gamma_{\mathrm{H} }} \Omega_{\mathrm{H}} \hat{\mathrm{H}}_{\mathrm{z}} + \Omega_{\mathrm{C}} \hat{\mathrm{C}}_{\mathrm{z}} \right) \right] 
\end{equation}
with $\gamma_i$ the gyromagnetic ratio of nucleus $i$ of offset $\Omega_i$, $t_1$ the increment in the indirect dimension and $in_1$ the increment giving $t=in_1 t_1$. At the end of the evolution at low field (Fig.\,\ref{fig:SpectrumPP}.a), ZQ coherences are selected by phase cycling. Importantly, the selected ZQ terms have evolved half of the time as ZQ and half of the time as DQ coherences to simultaneously scale the \Proton~and \Carbon~offsets and suppress the evolution due to spin-spin scalar couplings. It allows the correlation between ZQ \Carbon-\Proton~coherence at 0.33\,T and single quantum proton coherence at 14.1\,T on \ubiquitin~(Fig.\,\ref{fig:SpectrumPP}.b).\cite{cousin_recovering_2016} \\
In order to confirm our treatment of methyl group relaxation during multiple-quantum evolution, we simulated the spin evolution (relaxation, chemical shift and scalar coupling) at the low magnetic field center ($B_0$=0.33\,T) and calculated the corresponding spectrum. We used a spectral density function that accounts for the global tumbling of the protein (correlation time $\TAUC$), the methyl group rotation (order parameter $S_m^2$ and correlation time $\TAUM$), and two modes of motion of the C-C bond within the protein: fast (order parameter $S_f^2$ and correlation time $\TAUF$) and slow (order parameter $S_s^2$ and correlation time $\TAUS$). The spectral density function is written as:
\begin{equation}
	\begin{aligned}
		\mathcal{J}_{Ubi}(\omega, \theta_{\vec{\imath},\vec{\jmath}}) & = 
        \cfrac{2}{5}  [ S_m^2 (\theta_{\vec{\imath},\vec{\jmath}}) (S_f^2 S_s^2 L(\omega, \TAUC)    +  (1-S_f^2) L(\omega, \TAUF') + S_f^2(1-S_s^2) L(\omega, \TAUS') ) \\ 
        & + (\mathcal{P}_2 [cos(\theta_{\vec{\imath},\vec{\jmath}})] - S_m^2 (\theta_{\vec{\imath},\vec{\jmath}})) ( S_f^2 S_s^2 L(\omega, \TAUM')   + (1-S_f^2) L(\omega, \TAUF'') + S_f^2 (1-S_s^2) L(\omega, \TAUS''))]
    \end{aligned}
    \label{eq:SpectralDensityUbi}
\end{equation}
using the same definition as above. The values of the motional parameters for the isoleucine-$\delta_1$ methyl groups of ubiquitin used in the simulation of the spectrum were obtained in an independant study \cite{Cousin_JACS_2018} based on high-resolution NMR relaxometry of \ubiquitinRelax~and are reported in Supplementary Material Table\,S1.\\
Cross-section of the 2F-HZQC along the carbon dimension and simulated spectra are shown in Fig.\,\ref{fig:HZQCsimu}.a and Supplementary Material Fig.\,S5. Using a single relaxation rate $R_\mathrm{MQ,central}^{\mathrm{gen}}$ (Eq.\,\ref{eq:RNoAssump}) cannot explain the relatively sharp peaks at low field but is in perfect agreement with the HZQC spectrum recorded at a single high field (B$_0$=14.1\,T) (Fig.\,\ref{fig:HZQCsimu}.b and Supplementary Material Fig.\,S6). Our model, which considers the individual relaxation rates of the two contributions to the central line, reproduces well the linewidth of the peaks in both spectra. The low intensity observed for Ile-23 (Fig.\,\ref{fig:SpectrumPP}.b and supplementary material Fig.\,S6) may be due to slower rotation of the methyl group (Supplementary Material Table\,S1) as would be expected from Fig.\,\ref{fig:TmEffect}.

%% file: Section/7_Conclusion.tex
\section{Conclusion}
A general analysis of the relaxation properties of zero- and double-quantum coherences in methyl groups has been described, without invoking two key hypotheses of the original methyl-TROSY work:\cite{tugarinov_cross-correlated_2003} slow tumbling and fast methyl rotation, which are appropriate for large macromolecules at high fields. Symmetry considerations show that the free evolutions of ZQ (or DQ) coherences occur in a subspace of dimension 4. A numerical analysis shows that one component of this subspace relaxes slowly at all magnetic fields where the Redfield treatment of relaxation is valid. At high field, two operators relax slowly and correspond to the central lines of the methyl triplet. Analytical calculations with our model are then equivalent to the conventional methyl-TROSY theory. At low field, that is, where the slow tumbling approximation is not valid anymore, a single component relaxes slowly, preserving the methyl-TROSY effect for a third of the polarization. A detailed analysis of the spectral density functions that describe relaxation properties of the multiple-quantum coherences confirmed that the TROSY effect is optimal only under fast rotation of the methyl group from the pico- to few tens of pico-second time scales. This limits optimal TROSY conditions to un-constrained methyl groups with a free rotation around the symmetry axis. This particularity may hinder the observation of some conformationally constrained methyl-groups. Our comprenhensive approach shows that CSA/DD cross-correlated relaxation leads to more favorable relaxation properties for one component of the outer lines of the triplet at very high fields. At these magnetic fields, recording a HZQC using the cancellation effect brought by CSA/DD cross-correlation will lead to a new type of methyl TROSY. This new development sheds light on the 2F-HZQC experiment performed on \ubiquitin. It shows that the manipulation of ZQ coherences can be used to observe methyl groups of large macro-molecules at low magnetic fields where contributions of chemical exchange line broadening are dramatically reduced.\cite{cousin_recovering_2016}

%% file: Section/SI.tex
\beginsupplement
\subsection{Expressions of relaxation rates in the zero-quantum subspace}
\subsubsection{Auto-relaxation rates}
\begin{equation*}
	\begin{aligned}
    R(ZQ_\mathrm{outer,1}^\mathrm{A}) =& \frac{1}{8} d_{\sCH}^2 \left[8 \JCH(0) + 9 \JCH(\OMc) + 4 \JCH(\OMc-\OMh) + 9 \JCH(\OMh) + 12 \JCH(\OMc+\OMh)  \right] \\
   &+ \frac{1}{4} d_{\sCH}^2 \left[ 4 \JHCH(0) + 3 \JHCH(\OMc) + \JHCH(\OMc-\OMh) + 3 \JHCH(\OMh) + 6 \JHCH(\OMc+\OMh) \right] \\
    & - \frac{3}{2} d_{\sCH}d_{\sHH} \left[ 2 \JCHH(0) + \JCHH(\OMh) + 2 \JCHHH(0) - \JCHHH(\OMh) \right] \\ 
    &+ \frac{3}{4} d_{\sHH}^2 \left[ 3 \JHH(0) + 4 \JHH(\OMh) + 4 \JHH(2 \OMh) \right] \\
     &+ \frac{3}{4} d_{\sHH}^2 \left[ 3 \JHHH(0) + 2 \JHHH(\OMh) + 2 \JHHH(2 \OMh) \right]  \\
    &+ \frac{1}{18} \sigma_C^2 \OMc^2\left[ 4 \JCC(0) + 3 \JCC(\OMc) \right] \\
   &+ \frac{1}{3} d_{\sCH}  \sigma_C \OMc \left[ 4 \JCCH(0) + 3 \JCCH(\OMc) \right] - 2 d_{\sHH} \sigma_C \OMc \JCCHH(0) \\
    R(ZQ_\mathrm{outer,2}^\mathrm{A}) = & \frac{1}{8} d_{\sCH}^2 \left[ 8 \JCH(0) + 9 \JCH(\OMc) + 4 \JCH(\OMc-\OMh) + 9 \JCH(\OMh) + 12 \JCH(\OMc+\OMh) \right] \\
    &+ \frac{1}{4} d_{\sCH}^2 \left[ 4 \JHCH(0) + 3 \JHCH(\OMc) + \JHCH(\OMc-\OMh) + 3 \JHCH(\OMh) + 6 \JHCH(\OMc+\OMh) \right] \\
    & - \frac{3}{2} d_{\sCH}d_{\sHH} \left[ 2 \JCHH(0) + \JCHH(\OMh) + 2 \JCHHH(0) - \JCHHH(\OMh) \right] \\ 
    &+ \frac{3}{4} d_{\sHH}^2 \left[ 3 \JHH(0) + 4 \JHH(\OMh) + 4 \JHH(2 \OMh) \right] \\
    &+ \frac{3}{4} d_{\sHH}^2 \left[ 3 \JHHH(0) + 2 \JHHH(\OMh) + 2 \JHHH(2 \OMh) \right] \\
    &+ \frac{1}{18} \sigma_C^2\OMc^2 \left[ 4 \JCC(0) + 3 \JCC(\OMc) \right] \\
    &- \frac{1}{3} d_{\sCH} \sigma_C \OMc \left[ 4 \JCCH(0) + 3 \JCCH(\OMc) \right] + 2 d_{\sHH} \sigma_C \OMc \JCCHH(0)
    \end{aligned}
\end{equation*}

\begin{equation*}
	\begin{aligned}
    R(ZQ_\mathrm{central}^\mathrm{A}) = &\frac{1}{24} d_{\sCH}^2 \left[ 16 \JCH(0) + 27 \JCH(\OMc) + 12 \JCH(\OMc-\OMh) + 27 \JCH(\OMh) + 36 \JCH(\OMc+\OMh) \right]  \\
     &+ \frac{1}{12} d_{\sCH}^2 \left[ - 8 \JHCH(0) - 9 \JHCH(\OMc) + 6 \JHCH(\OMc-\OMh) + 18 \JHCH(\OMh) + 36 \JHCH(\OMc+\OMh) \right] \\
    &- \frac{1}{2} d_{\sCH} d_{\sHH} \left[  4 \JCHH(0) + 3 \JCHH(\OMh) \right] + d_{\sCH} d_{\sHH} \left[ 2 \JCHHH(0) - 3 \JCHHH(\OMh)  \right]\\
    &+ \frac{3}{4} d_{\sHH}^2 \left[ 2 \JHH(0) + 5 \JHH(\OMh) + 2 \JHH(2 \OMh) \right] \\
    &+ \frac{3}{4} d_{\sHH}^2 \left[ - 2 \JHHH(0) + \JHHH(\OMh) + 4 \JHHH(2 \OMh) \right] \\
    &+ \frac{1}{18} \sigma_C^2 \OMc^2 \left[ 4 \JCC(0) + 3 \JCC(\OMc) \right] \\
    R(ZQ_{\mathrm{central}}^{\mathrm{\Sigma E}}) = & \frac{1}{24} d_{\sCH}^2 \left[ 8 \JCH(0) + 27 \JCH(\OMc) + 12 \JCH(\OMc-\OMh) + 27 \JCH(\OMh) + 36 \JCH(\OMc+\OMh) \right] \\
    &- \frac{1}{12} d_{\sCH}^2 \left[ 4 \JHCH(0) + 9 \JHCH(\OMc) + 3 \JHCH(\OMc-\OMh) + 9 \JHCH(\OMh) + 18 \JHCH(\OMc+\OMh) \right] \\
   & - \frac{1}{2} d_{\sCH} d_{\sHH} \left[ 2 \JCHH(0) + 3 \JCHH(\OMh) - 2 \JCHHH(0) - 3 \JCHHH(\OMh) \right] \\
   &+ \frac{3}{4} d_{\sHH}^2 \left[ \JHH(0) + 2 \JHH(\OMh) + 2 \JHH(2 \OMh) \right] \\
   &- \frac{3}{4} d_{\sHH}^2 \left[\JHHH(0) + 2 \JHHH(\OMh) + 2 \JHHH(2 \OMh) \right] \\
   &+ \frac{1}{18} \sigma_C^2 \OMc^2 \left[ 4 \JCC(0) + 3 \JCC(\OMc) \right]
    \end{aligned}
\end{equation*}

\subsubsection{Cross-relaxation rates}

\begin{equation*}
	\begin{aligned}
    R(ZQ_{\mathrm{outer},1}^\mathrm{A} \leftrightarrow ZQ_{\mathrm{outer},2}^\mathrm{A}) = & -\frac{3}{2} d_{\sHH}^2 \JHH(2 \OMh) - 3 d_{\sHH}^2 \JHHH(2 \OMh) \\
    R(ZQ_{\mathrm{outer},1}^\mathrm{A} \leftrightarrow ZQ_{\mathrm{central}}^\mathrm{A}) = &  \frac{\sqrt{3}}{4} d_{\sCH}^2 \JCH(\OMh) + \frac{\sqrt{3}}{2} d_{\sCH}^2 \JHCH(\OMh) - \frac{\sqrt{3}}{2} d_{\sCH} d_{\sHH}\left[2 \JCHH(\OMh) + \JCHHH(\OMh) \right] \\
    R(ZQ_{\mathrm{outer},1}^\mathrm{A} \leftrightarrow ZQ_{\mathrm{central}}^{\mathrm{\Sigma E}}) = & \frac{3}{4 \sqrt{6}} d_{\sCH}^2 \left[\JCH(\OMh) - \JHCH(\OMh) \right] + \frac{9}{4 \sqrt{6}} d_{\sHH}^2 \left[ \JHH(\OMh) - \JHHH(\OMh) \right] \\
   & - \frac{3}{\sqrt{6}} d_{\sCH}d_{\sHH} \left[ \JCHH(\OMh) - \JCHHH(\OMh) \right] \\
    R(ZQ_{\mathrm{outer},2}^\mathrm{A} \leftrightarrow ZQ_{\mathrm{central}}^\mathrm{A}) = & R(ZQ_{\mathrm{outer},1}^\mathrm{A} \leftrightarrow ZQ_{\mathrm{central}}^\mathrm{A}) \\
    R(ZQ_{\mathrm{outer},2}^\mathrm{A} \leftrightarrow ZQ_{\mathrm{central}}^{\mathrm{\Sigma E}}) = & R(ZQ_{\mathrm{outer},1}^\mathrm{A} \leftrightarrow ZQ_{\mathrm{central}}^{\mathrm{\Sigma E}}) \\
    R(ZQ_{\mathrm{central}}^\mathrm{A} \leftrightarrow ZQ_{\mathrm{central}}^{\mathrm{\Sigma E}}) = & \frac{\sqrt{2}}{3}   d_{\sCH}^2 \left[ \JCH(0) - \JHCH(0) \right] + \frac{3}{4} \sqrt{2} d_{\sHH}^2 \left[ \JHH(0) - \JHHH(0) \right]  \\
	&+ \sqrt{2} d_{\sCH} d_{\sHH} \left[ \JCHHH(0) - \JCHH(0) \right]
	\end{aligned}
\end{equation*}

\newpage

\subsection{Expressions of relaxation rates in the double-quantum subspace}
\subsubsection{Auto-relaxation rates}

\begin{equation*}
	\begin{aligned}
    R(DQ_{\mathrm{outer},1}^\mathrm{A}) = & \frac{1}{8} d_{\sCH}^2 \left[ 8 \JCH(0) + 9 \JCH(\OMc) + 2 \JCH(\OMc-\OMh) + 9 \JCH(\OMh) + 24 \JCH(\OMc+\OMh) 
\right] \\
	& + \frac{1}{4} d_{\sCH}^2 \left[ 4 \JHCH(0) + 3 \JHCH(\OMc) + \JHCH(\OMc-\OMh) + 3 \JHCH(\OMh) + 6 \JHCH(\OMc+\OMh) \right] \\
         & + \frac{3}{2} d_{\sCH}d_{\sHH} \left[ 2 \JCHH(0) + \JCHH(\OMh) + 2 \JCHHH(0) - \JCHHH(\OMh) \right] \\
	& + \frac{3}{4} d_{\sHH}^2 \left[ 3 \JHH(0) + 4 \JHH(\OMh) + 4 \JHH(2 \OMh) \right] \\
	& + \frac{3}{4} d_{\sHH}^2 \left[3 \JHHH(0) + 2 \JHHH(\OMh) + 2 \JHHH(2 \OMh) \right] \\
	& + \frac{1}{18} \sigma_C^2 \OMc^2 \left[ 4 \JCC(0) + 3 \JCC(\OMc) \right] \\
	& + \frac{1}{3} d_{\sCH} \sigma_C \OMc \left[4 \JCCH(0) + 3 \JCCH(\OMc) \right] +2 d_{\sHH} \sigma_C \OMc \JCCHH (0) \\
    R(DQ_{\mathrm{outer},2}^\mathrm{A}) = &  \frac{1}{8} d_{\sCH}^2 \left[ 8 \JCH(0) + 9 \JCH(\OMc) + 2 \JCH(\OMc-\OMh) + 9 \JCH(\OMh) + 24 \JCH(\OMc+\OMh) 
\right] \\
	& + \frac{1}{4} d_{\sCH}^2 \left[ 4 \JHCH(0) + 3 \JHCH(\OMc) + \JHCH(\OMc-\OMh) + 3 \JHCH(\OMh) + 6 \JHCH(\OMc+\OMh) \right] \\
         & + \frac{3}{2} d_{\sCH}d_{\sHH} \left[ 2 \JCHH(0) + \JCHH(\OMh) + 2 \JCHHH(0) - \JCHHH(\OMh) \right] \\
	& + \frac{3}{4} d_{\sHH}^2 \left[ 3 \JHH(0) + 4 \JHH(\OMh) + 4 \JHH(2 \OMh) \right] \\
	& + \frac{3}{4} d_{\sHH}^2 \left[3 \JHHH(0) + 2 \JHHH(\OMh) + 2 \JHHH(2 \OMh) \right] \\
	& + \frac{1}{18} \sigma_C^2 \OMc^2 \left[ 4 \JCC(0) + 3 \JCC(\OMc) \right] \\
	& - \frac{1}{3} d_{\sCH} \sigma_C \OMc \left[4 \JCCH(0) + 3 \JCCH(\OMc) \right] - 2 d_{\sHH} \sigma_C \OMc \JCCHH (0) \\
    R(DQ_{\mathrm{central}}^\mathrm{A}) = &  \frac{1}{24} d_{\sCH}^2 \left[ 16 \JCH(0) + 27 \JCH(\OMc) + 6 \JCH(\OMc-\OMh) + 27 \JCH(\OMh) + 72 \JCH(\OMc+\OMh) 
\right] \\
	& + \frac{1}{12} d_{\sCH}^2 \left[ - 8 \JHCH(0) - 9 \JHCH(\OMc) + 6 \JHCH(\OMc-\OMh) + 18 \JHCH(\OMh) + 36 \JHCH(\OMc+\OMh) \right] \\
         & + \frac{1}{2} d_{\sCH} d_{\sHH} \left[ 4 \JCHH(0) + 3 \JCHH(\OMh) \right] - d_{\sCH}d_{\sHH} \left[ 2 \JCHHH(0) - 3 \JCHHH(\OMh) \right] \\
	& + \frac{3}{4} d_{\sHH}^2 \left[ 2 \JHH(0) + 5 \JHH(\OMh) + 2 \JHH (2 \OMh) \right] \\
	& + \frac{3}{4} d_{\sHH}^2 \left[ - 2 \JHHH(0) + \JHHH(\OMh) + 4 \JHHH(2 \OMh) \right] \\
	& + \frac{1}{18} \sigma_C^2 \OMc^2 \left[ 4 \JCC(0) + 3 \JCC(\OMc) \right] \\
	R(DQ_{\mathrm{central}}^{\mathrm{\Sigma E}}) = & \frac{1}{24} d_{\sCH}^2 \left[ 8 \JCH(0) + 27 \JCH(\OMc) + 6 \JCH(\OMc-\OMh) + 27 \JCH(\OMh) + 72 \JCH(\OMc+\OMh) \right]\\
	& - \frac{1}{12} d_{\sCH}^2 \left[ 4 \JHCH(0) + 9 \JHCH(\OMc) + 3 \JHCH(\OMc-\OMh) + 9 \JHCH(\OMh) + 18 \JHCH(\OMc+\OMh) \right] \\
         & + \frac{1}{2} d_{\sCH} d_{\sHH} \left[ 2 \JCHH(0) + 3 \JCHH(\OMh) - 2 \JCHHH(0) - 3 \JCHHH(\OMh) \right] \\
	& + \frac{3}{4} d_{\sHH}^2 \left[ \JHH(0) + 2 \JHH(\OMh) + 2 \JHH(2 \OMh) \right] \\
	& - \frac{3}{4} d_{\sHH}^2 \left[ \JHHH(0) + 2 \JHHH(\OMh) + 2 \JHHH(2 \OMh) \right] \\
	& + \frac{1}{18} \sigma_C^2 \OMc^2 \left[ 4 \JCC(0) + 3 \JCC(\OMc) \right]
	\end{aligned}
\end{equation*}

\subsubsection{Cross-relaxation rates}

\begin{equation*}
	\begin{aligned}
    R(DQ_{\mathrm{outer},1}^\mathrm{A} \leftrightarrow DQ_{\mathrm{outer},2}^\mathrm{A}) = & R(ZQ_{\mathrm{outer},1}^\mathrm{A} \leftrightarrow ZQ_{\mathrm{outer},2}^\mathrm{A}) \\
    R(DQ_{\mathrm{outer},1}^\mathrm{A} \leftrightarrow DQ_{\mathrm{central}}^\mathrm{A}) = &  \frac{\sqrt{3}}{4} d_{\sCH}^2 \JCH(\OMh) + \frac{\sqrt{3}}{2} d_{\sCH}^2 \JHCH(\OMh) + \frac{\sqrt{3}}{2} d_{\sCH} d_{\sHH} \left[ 2 \JCHH(\OMh) + \JCHHH(\OMh) \right] \\
    R(DQ_{\mathrm{outer},1}^\mathrm{A} \leftrightarrow DQ_{\mathrm{central}}^{\mathrm{\Sigma E}}) = & \frac{3}{4 \sqrt{6}} d_{\sCH}^2 \left[\JCH(\OMh) - \JHCH(\OMh) \right] + \frac{9}{4 \sqrt{6}} d_{\sHH}^2 \left[ \JHH(\OMh) - \JHHH(\OMh) \right] \\
   & + \frac{3}{\sqrt{6}} d_{\sCH}d_{\sHH} \left[ \JCHH(\OMh) - \JCHHH(\OMh) \right] \\
    R(DQ_{\mathrm{outer},2}^\mathrm{A} \leftrightarrow DQ_{\mathrm{central}}^\mathrm{A}) = & R(DQ_{\mathrm{outer},1}^\mathrm{A} \leftrightarrow DQ_{\mathrm{central}}^\mathrm{A}) \\
    R(DQ_{\mathrm{outer},2}^\mathrm{A} \leftrightarrow DQ_{\mathrm{central}}^{\mathrm{\Sigma E}}) = & R(DQ_{\mathrm{outer},1}^\mathrm{A} \leftrightarrow DQ_{\mathrm{central}}^{\mathrm{\Sigma E}}) \\
    R(DQ_{\mathrm{central}}^\mathrm{A} \leftrightarrow DQ_{\mathrm{central}}^{\mathrm{\Sigma E}}) =  & \frac{\sqrt{2}}{3}   d_{\sCH}^2 \left[ \JCH(0) - \JHCH(0) \right] +\frac{3}{4} \sqrt{2} d_{\sHH}^2 \left[ \JHH(0) - \JHHH(0) \right]  \\
	&- \sqrt{2} d_{\sCH} d_{\sHH} \left[ \JCHHH(0) - \JCHH(0) \right]
	\end{aligned}
\end{equation*}

\newpage

\subsection{Dynamics parameters used for \ubiquitin}
\begin{table}[!ht]
	\caption{Order parameters and correlation times for \ubiquitin~previously determined using high-resolution relaxometry on \ubiquitinRelax.\cite{Cousin_JACS_2018} $\TAUC$ was set to 5.028\,ns.}
	\begin{center}
		{\def\arraystretch{1.3}
		\begin{tabular}{ccccccc}
			Residue & $S_f^2$  & $S_s^2$ & $\TAUM$ (ps) & $\TAUF$ (ps) & $\TAUS$ (ns) & CSA (ppm) \\
			\hline
		 	3 & 0.71  & 0.76  & 8.56 & 34.8 & 17.7 & 23.75 \\%
		 	13 & 0.63  & 0.62 & 11.5 & 77.3 & 3.13  & 23.88 \\%
			23 & 0.51  & 0.90  & 21.7 & 150 & 18.3  & 20.14 \\%
			30 & 0.81  & 0.74  & 9.03 & 61.8 & 6.98 & 22.53 \\%
			36 & 0.67  & 0.58 & 8.05 & 82.8 & 2.48  & 29.37 \\%
			44 & 0.51  & 0.28 & 5.58 & 70.3 & 1.27  & $25.0\footnotemark[1]$ \\%
			61 & 0.55  & 0.89 & 14.4 & 138 & 17.3  & 26.11 \\%
		\end{tabular}
		}
	\end{center}
	\label{Table:DynParam}
	 \footnotetext[1]{Chemical exchange is affecting transverse relaxation rates of Ile-44 and CSA was kept constant at 25.0\,ppm.}
\end{table}

\newpage

\subsection{Figures}

\begin{figure*}[!ht]
	\begin{center}
		\includegraphics[width=0.8\textwidth]{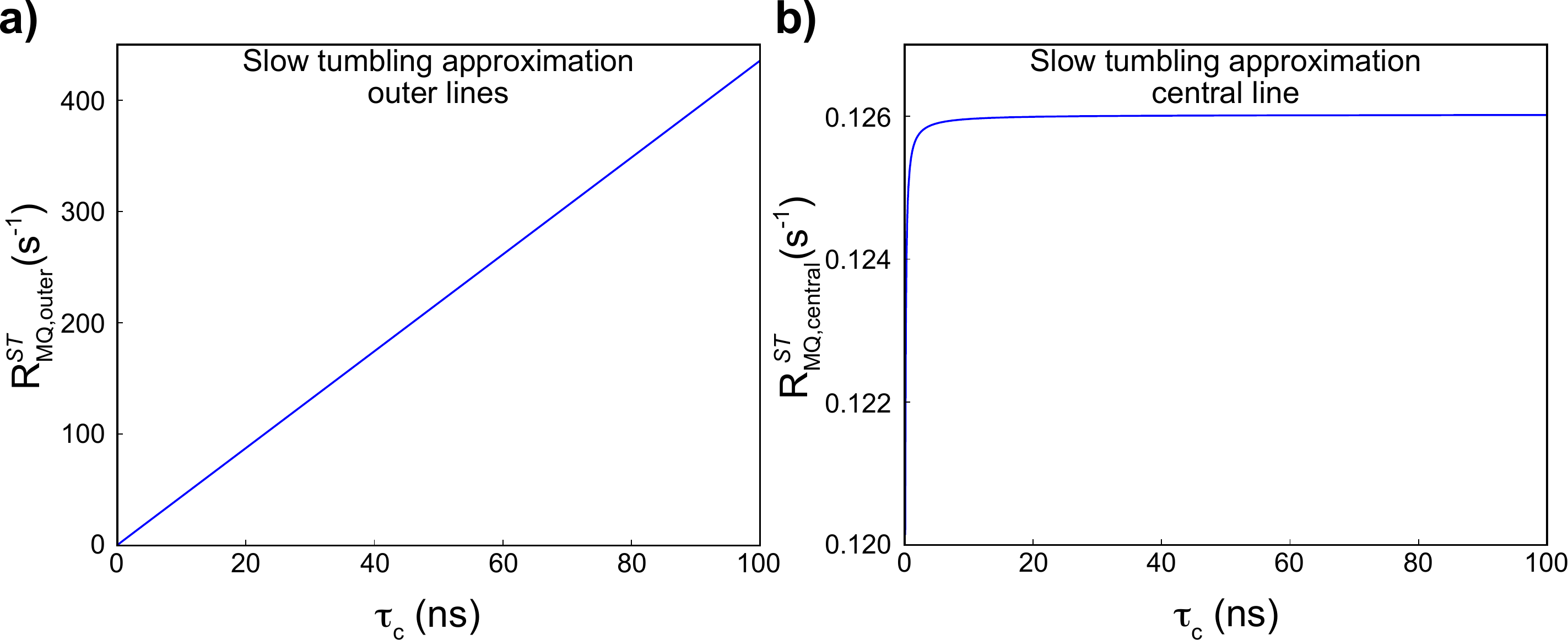}
	\end{center}
	\caption{Variation of the DD contributions to auto-relaxation rates in the \SlowTumbling~approximation with the correlation time $\TAUC$ for the outer lines (\textbf{a}) and the central line (\textbf{b}). Relaxation rates were calculated using the spectral density function $\mathcal{J}_{MFM}$. The non-zero rates predicted for the central line are due to finite speed of methyl rotation.}
	\label{fig:EvolutionRatesVsTCSI}
\end{figure*}

\begin{figure*}[!ht]
	\begin{center}
		\includegraphics[width=0.8\textwidth]{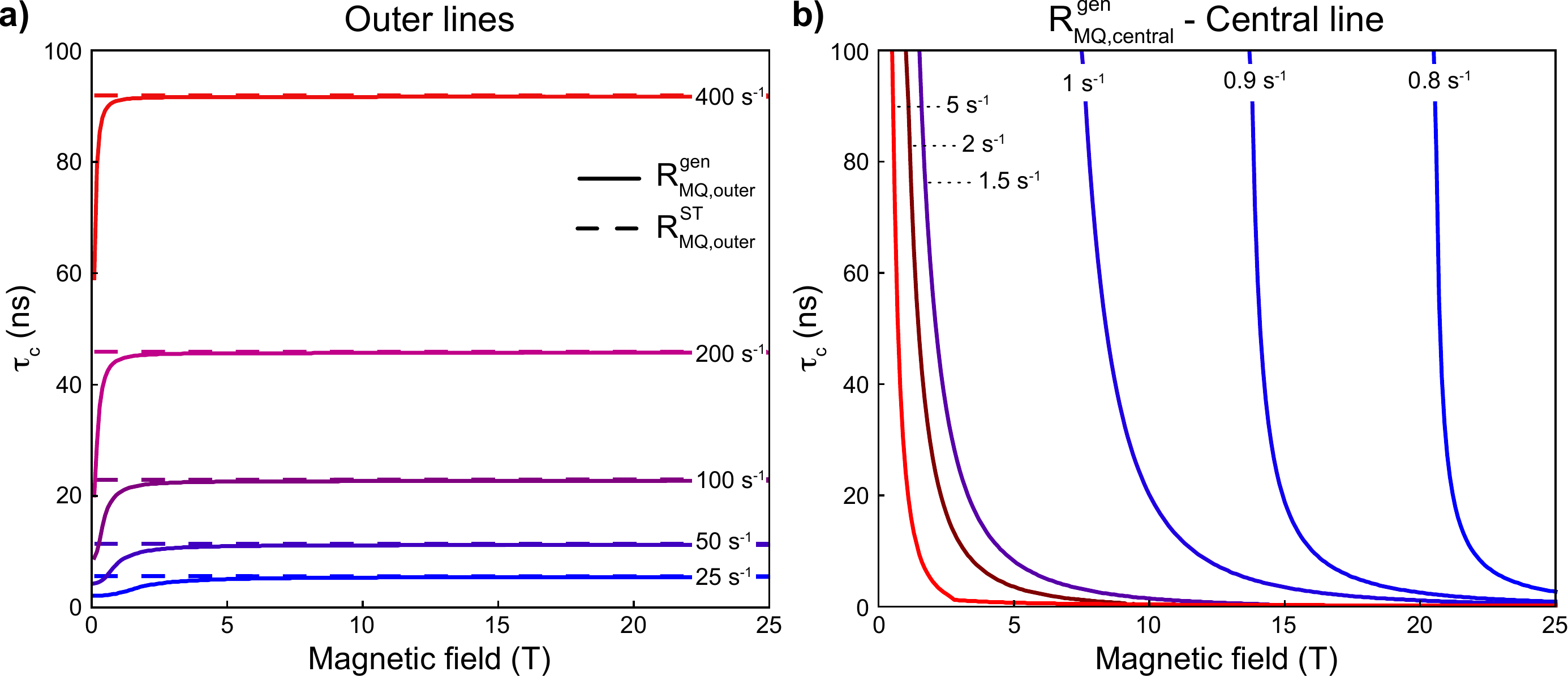}
	\end{center}
	\caption{Domain of validity of the \SlowTumbling~regime. \textbf{a.} Contour plot of the relaxation rate of the outer lines of the triplet as a function of the magnetic field and the correlation time $\TAUC$ and calculated using Eq.\,\ref{eq:RNoAssump} ($R_{\mathrm{MQ,outer}}^{\mathrm{gen}}$, solid line) and Eq.\,\ref{eq:ExtremeNarrowing} ($R_{\mathrm{MQ,outer}}^{ST}$, dashed line). The \SlowTumbling~approximation holds when the dashed and solid lines coincide. \textbf{b.}  Contour plot of the relaxation rate of the central line of the triplet as a function of the magnetic field and the correlation time $\TAUC$ and calculated using Eq.\,\ref{eq:RNoAssump} ($R_{\mathrm{MQ,central}}^{\mathrm{gen}}$). By comparison, the expected relaxation rate for the central single-transition operator $T_\mathrm{central}$ in the \SlowTumbling~approximation is $R_{\mathrm{MQ,central}}^{ST} \approx R_{\mathrm{MQ,central}}^{ST,IFR}=0$ (Supplementary Material Fig.\,\ref{fig:EvolutionRatesVsTCSI}.a).}
	\label{fig:LimitST}
\end{figure*}

\begin{figure*}[!ht]
	\begin{center}
		\includegraphics[width=0.8\textwidth]{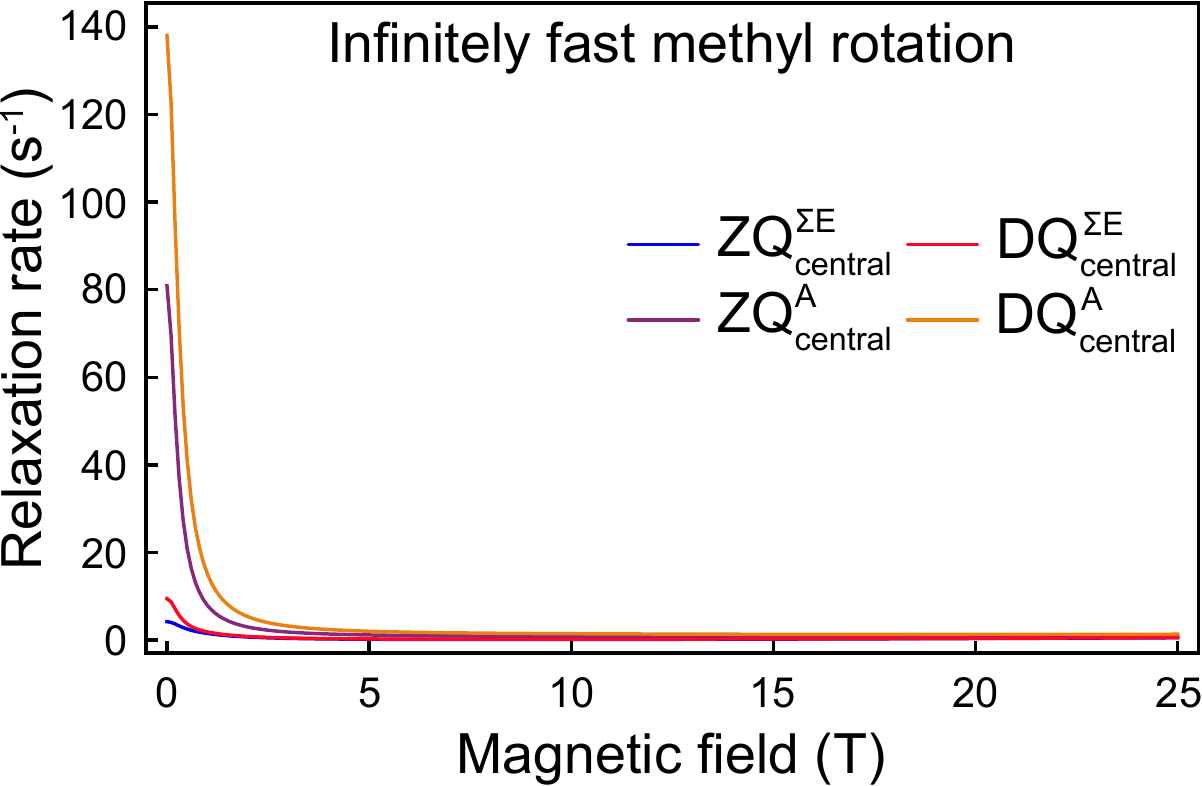}
	\end{center}
	\caption{Auto-relaxation rates of the operators $ZQ_{\mathrm{central}}^\mathrm{\Sigma E}$, $ZQ_{\mathrm{central}}^\mathrm{A}$, $DQ_{\mathrm{central}}^\mathrm{\Sigma E}$ and $DQ_{\mathrm{central}}^\mathrm{A}$ from 0.1 to 25\,T. Relaxation rates were calculated using the spectral density function $\mathcal{J}_{MFM}^{IFR}$ and the same dynamic parameters as detailed in the main text.}
	\label{fig:IFRcentralOperators}
\end{figure*}

\begin{figure*}[!ht]
	\begin{center}
		\includegraphics[width=0.8\textwidth]{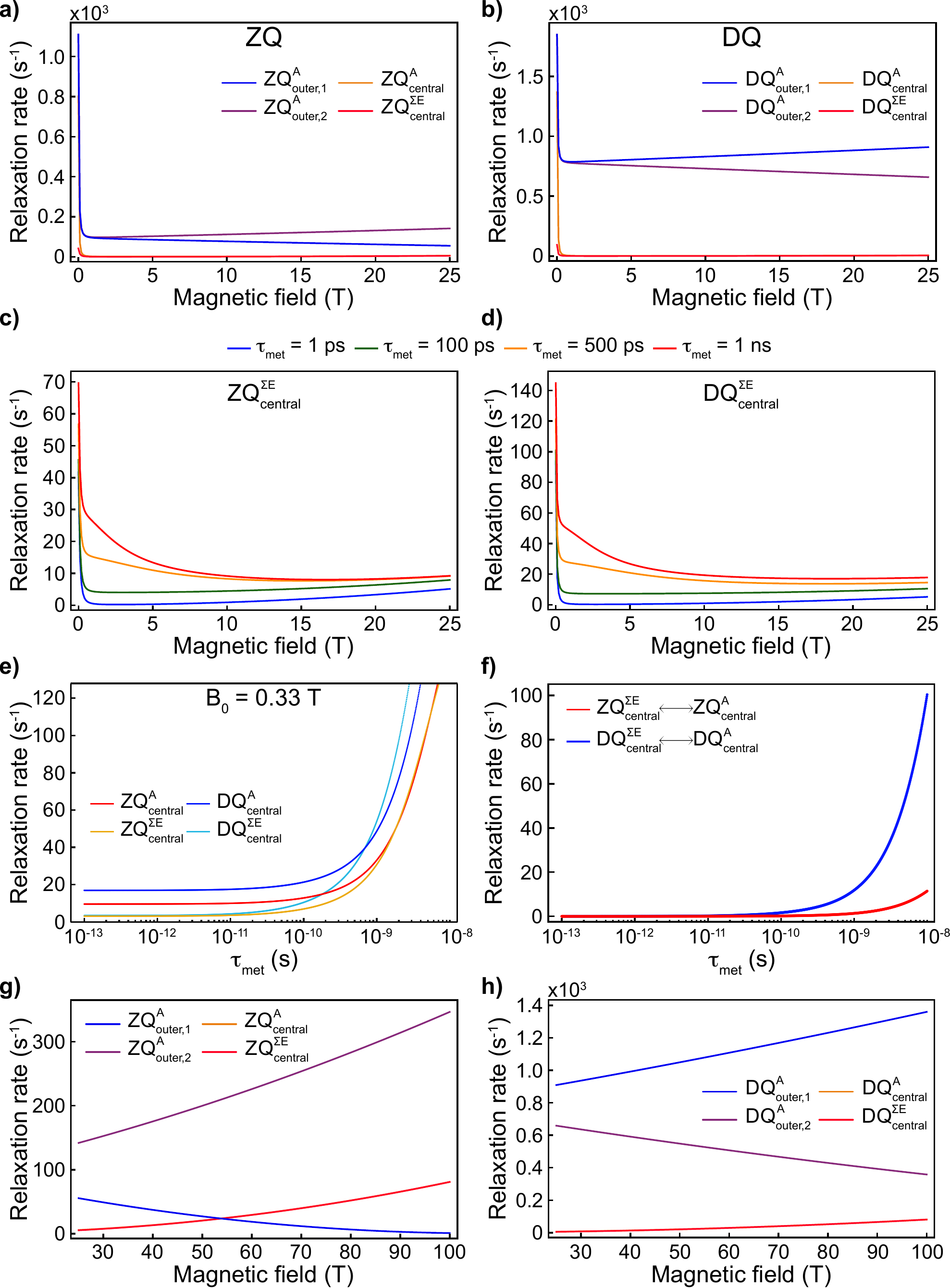}
	\end{center}
	\caption{Methyl-TROSY effect in the case of a large protein ($\TAUC$ = 100\,ns). Auto-relaxation rates in the bases (\textbf{a}) $\mathcal{B}_{ZQ}$ and (\textbf{b}) $\mathcal{B}_{DQ}$. Evolution of the auto-relaxation rates of the operators (\textbf{c}) $ZQ_{\mathrm{central}}^\mathrm{\Sigma E}$ and (\textbf{d}) $DQ_{\mathrm{central}}^\mathrm{\Sigma E}$ from 0.1 to 25\,T with $\TAUM$ values ranging from 1\,ps to 1\,ns. \textbf{e.} Evolution of the auto-relaxation rates of the operators contributing to the central line of the triplet as a function of the methyl-group rotational correlation time $\TAUM$ at 0.33\,T. \textbf{f.} Evolution of the intra-line cross-relaxation rates as a function of the methyl-group rotational correlation time $\TAUM$. Magnetic-field dependence from 25 to 100 T of the auto-relaxation rates of the coherences contributing to the central and outer lines of the triplet in the bases (\textbf{g}) $\mathcal{B}_{ZQ}$ and (\textbf{h}) $\mathcal{B}_{DQ}$. Relaxation rates were calculated using the spectral density function $\mathcal{J}_{MFM}$ and the same dynamic parameters as detailed in the main text, except for $\TAUC$ = 100\,ns.}
	\label{fig:HighMW}
\end{figure*}

\begin{figure*}[!ht]
	\begin{center}
		\includegraphics[width=0.75\textwidth]{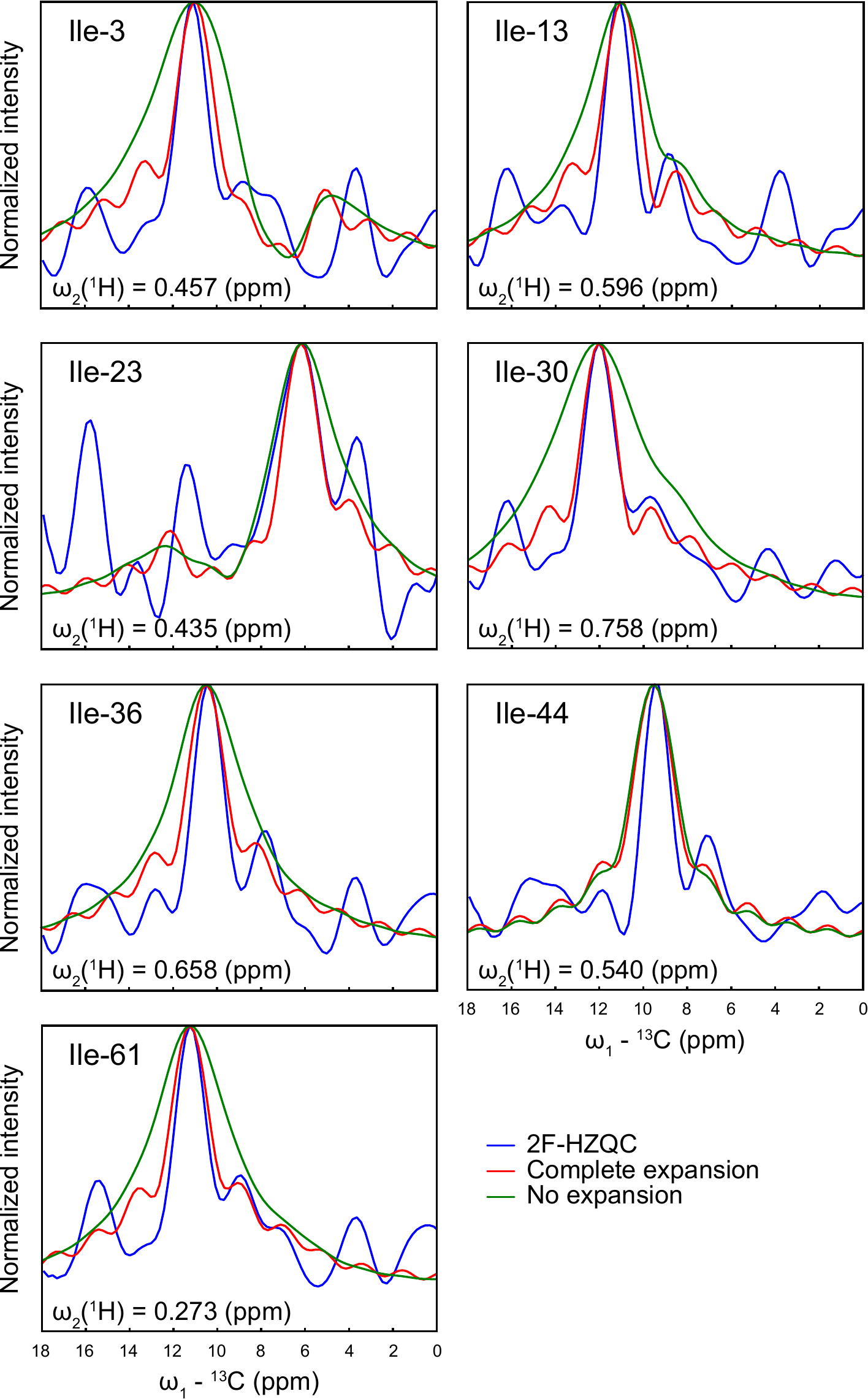}
	\end{center}
	\caption{Simulation of the 2F-HZQC spectrum at 0.33\,T of \ubiquitin. Simulation of cross-sections along the indirect dimension of the 2F-HZQC spectrum for the seven isoleucine residues of \ubiquitin~measured with $t_1$ evolution at 0.33\,T. The cross-sections from the experimental spectra (blue) are compared with a simulation using our approach (red) or using the previously reported expression of relaxation rates for the central lines of the triplets (green). All cross-sections are normalized independently so that all spectra have the same maximum intensity.}
	\label{fig:HZQCsimuSI}
\end{figure*}

\begin{figure*}[!ht]
	\begin{center}
		\includegraphics[width=0.8\textwidth]{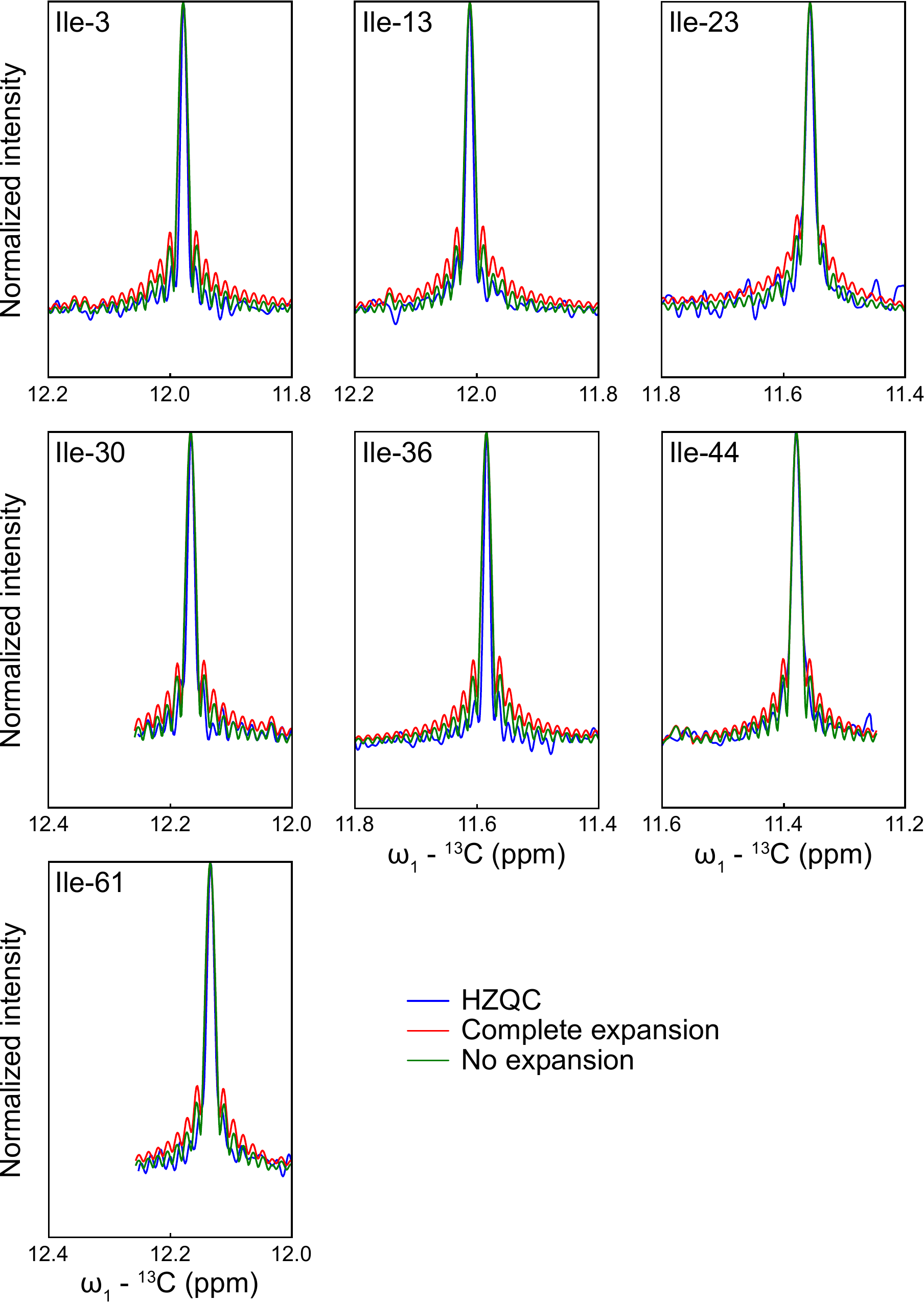}
	\end{center}
	\caption{Simulation of the HZQC spectrum at 14.1\,T of \ubiquitin. Simulation of cross-sections along the indirect dimension of the 2F-HZQC spectrum for the seven isoleucine residues of \ubiquitin~measured with $t_1$ evolution at 14.1\,T. The cross-sections from the experimental spectra (blue) are compared with simulations using our approach (red) or using the previously reported expression of relaxation rates for the central lines of the triplets (green). All cross-sections are normalized independently so that all spectra have the same maximum intensity.}
	\label{fig:HFHZQCsimu}
\end{figure*}

%% file: Article.bbl
\providecommand{\noopsort}[1]{}\providecommand{\singleletter}[1]{#1}
\begin{thebibliography}{10}

\bibitem{Cavanagh_ProteinNMR_2007}
J.~Cavanagh, W.~J. Fairbrother, A.~G. Palmer, M.~Rance, and N.~J. Skelton, {\em
  Protein NMR Spectroscopy: Principles and Practice}.
\newblock Elsevier Academic Press, 2007.

\bibitem{Palmer_ChemRev_2004}
A.~G. Palmer {\em Chem.\,Rev.}, vol.~104, p.~3640, 2004.

\bibitem{Mittermaier_science_2006}
A.~Mittermaier and L.~E. Kay {\em Science}, vol.~312, p.~224, 2006.

\bibitem{Charlier_ChemSocRev_2016}
C.~Charlier, S.~Cousin, and F.~Ferrage {\em Chem.\,Soc.\,Rev.}, vol.~45,
  p.~2410, 2016.

\bibitem{Ardenkjaer_Angewandte_2015}
J.-H. Ardenkjaer-Larsen, G.~S. Boebinger, A.~A. Comment, S.~Duckett, A.~S.
  Edison, F.~Engelke, C.~Griesinger, R.~G. Griffin, C.~Hilty, H.~Maeda,
  G.~Parigi, T.~Prisner, E.~Ravera, J.~van Bentum, S.~Vega, A.~Webb,
  C.~Luchinat, H.~Schwalbe, and L.~Frydman {\em Angew.\,Chem.\,Int.\,Ed.},
  vol.~54, p.~9162, 2015.

\bibitem{Pervushin_attenuated_1997}
K.~Pervushin, R.~Riek, G.~Wider, and K.~Wuthrich {\em
  Proc.\,Natl.\,Acad.\,Sci.\,USA}, vol.~94, p.~12366, 1997.

\bibitem{Shimizu_JCP_1964}
H.~Shimizu {\em J.\,Chem.\,Phys.}, vol.~40, p.~3357, 1964.

\bibitem{Goldman_JMagRes_1984}
M.~Goldman {\em J.\,Magn.\,Reson.}, vol.~60, p.~437, 1984.

\bibitem{Wimperis_MolPhys_1989}
S.~Wimperis and G.~Bodenhausen {\em Molec.\,Phys.}, vol.~66, p.~897, 1989.

\bibitem{Tugarinov_fourdimensional_2002}
V.~Tugarinov, R.~Muhandiram, A.~Ayed, and L.~E. Kay {\em J.\,Am.\,Chem.\,Soc.},
  vol.~124, p.~10025, 2002.

\bibitem{fiaux_nmr_2002}
J.~Fiaux, E.~B. Bertelsen, A.~L. Horwich, and K.~Wuthrich {\em Nature},
  vol.~418, p.~207, 2002.

\bibitem{hajduk_nmr-based_2000}
P.~J. Hajduk, D.~J. Augeri, J.~Mack, R.~Mendoza, J.~Yang, S.~F. Betz, and S.~W.
  Fesik {\em J.\,Am.\,Chem.\,Soc.}, vol.~122, p.~7898, 2000.

\bibitem{DuBay_AccChemRes_2015}
K.~H. DuBay, G.~R. Bowman, and P.~L. Geissler {\em Acc.\,Chem.\,Res.}, vol.~48,
  p.~1098, 2015.

\bibitem{Frederick_Nature_2007}
K.~K. Frederick, M.~S. Marlow, K.~G. Valentine, and J.~Wand {\em Nature},
  vol.~448, p.~325, 2007.

\bibitem{Nicholson_dynamics_1992}
L.~K. Nicholson, L.~E. Kay, D.~M. Baldisseri, J.~Arango, P.~E. Young, A.~Bax,
  and D.~A. Torchia {\em Biochemistry}, vol.~31, p.~5253, 1992.

\bibitem{Cousin_JACS_2018}
S.~F. Cousin, P.~Kaderavek, N.~Bolik-Coulon, Y.~Gu, C.~Charlier, L.~Carlier,
  L.~Bruschweiler-Li, T.~Marquardsen, J.-M. Tyburn, R.~Bruschweiler, and
  F.~Ferrage {\em J.\,Am.\,Chem.\,Soc.}, vol.~140, p.~13456, 2018.

\bibitem{mas_SciAdv_2018}
G.~Mas, J.-Y. Guan, E.~Crublet, E.~Colas~Debled, C.~Moriscot, P.~Gans,
  G.~Schoehn, P.~Macek, and J.~Boisbouvier {\em Sci.\,Adv.}, vol.~4, p.~1,
  2018.

\bibitem{tugarinov_cross-correlated_2003}
V.~Tugarinov, P.~M. Hwang, J.~E. Ollerenshaw, and L.~E. Kay {\em
  J.\,Am.\,Chem.\,Soc.}, vol.~125, p.~10420, 2003.

\bibitem{Sheppard_Progress_2010}
D.~Sheppard, R.~Sprangers, and V.~Tugarinov {\em Prog.\,Nucl.\,Magn.\,Reson.},
  vol.~56, p.~1, 2010.

\bibitem{tugarinov_isotope_2006}
V.~Tugarinov, V.~Kanelis, and L.~E. Kay {\em Nat.\,Protoc.}, vol.~1, p.~749,
  2006.

\bibitem{clore_chapter_2012}
M.~J. Plevin and J.~Boisbouvier, {\em Isotope-Labelling of Methyl Groups for
  NMR Studies of Large Proteins}, ch.~1, pp.~1--24.
\newblock Royal Society of Chemistry, 2012.

\bibitem{mas_specific_2013}
G.~Mas, E.~Crublet, O.~Hamelin, P.~Gans, and J.~Boisbouvier {\em
  J.\,Biomol.\,NMR}, vol.~57, p.~251, 2013.

\bibitem{rosenzweig_bringing_2014}
R.~Rosenzweig and L.~E. Kay {\em Annu.\,Rev.\,Biochem.}, vol.~83, p.~291, 2014.

\bibitem{tugarinov_JACS_2007}
V.~Tugarinov, R.~Sprangers, and L.~E. Kay {\em J.\,Am.\,Chem.\,Soc.}, vol.~129,
  p.~1743, 2007.

\bibitem{cousin_recovering_2016}
S.~F. Cousin, P.~Kade{\v r}{\'a}vek, B.~Haddou, C.~Charlier, T.~Marquardsen,
  J.-M. Tyburn, P.-A. Bovier, F.~Engelke, W.~Maas, G.~Bodenhausen,
  P.~Pelupessy, and F.~Ferrage {\em Angew.\,Chem.\,Int.\,Ed.}, vol.~55,
  p.~9886, 2016.

\bibitem{tugarinov_line_2004}
V.~Tugarinov, R.~Sprangers, and L.~E. Kay {\em J.\,Am.\,Chem.\,Soc.}, vol.~126,
  p.~4921, 2004.

\bibitem{cousin_high-resolution_2016}
S.~F. Cousin, C.~Charlier, P.~Kade{\v r}{\'a}vek, T.~Marquardsen, J.-M. Tyburn,
  P.-A. Bovier, S.~Ulzega, T.~Speck, D.~Wilhelm, F.~Engelke, W.~Maas,
  D.~Sakellariou, G.~Bodenhausen, P.~Pelupessy, and F.~Ferrage {\em
  Phys.\,Chem.\,Chem.\,Phys.}, vol.~18, p.~33187, 2016.

\bibitem{Blumich_ACIE_2017}
B.~Blumich and K.~Singh {\em Angew. Chem. Int. Ed.}, vol.~57, p.~6996, 2017.

\bibitem{Gan_JMR_2017}
Z.~Gan, I.~Hung, X.~Wang, J.~Paulino, W.~Gang, I.~M. Litvak, P.~L. Gor'kov,
  W.~W. Grey, P.~Lendi, J.~L. Schiano, M.~D. Bird, I.~R. Dixon, J.~Toth,
  G.~Boebinger, and T.~A. Cross {\em J.\,Magn.\,Reson.}, vol.~284, p.~125,
  2017.

\bibitem{wolfram_research_inc._mathematica_2016}
I.~Wolfram~Research, ``Mathematica,'' 2016.

\bibitem{Bengs_MagnResonChecm_2017}
C.~Bengs and M.~H. Levitt {\em Magn.\,Reson.\,Chem.}, vol.~56, p.~374, 2018.

\bibitem{kuprov_bloch-redfield-wangsness_2007}
I.~Kuprov, N.~Wagner-Rundell, and P.~Hore {\em J.\,Magn.\,Reson.}, vol.~184,
  p.~196, 2007.

\bibitem{Helmus_JbiolNMR_2013}
J.~J. Helmus and C.~P. Jaroniec {\em J.\,Biomol.\,NMR}, vol.~55, pp.~355--367,
  2013.

\bibitem{Werbelow_JCP_1975}
L.~G. Werbelow and D.~M. Grant {\em J.\,Chem.\,Phys.}, vol.~63, p.~544, 1975.

\bibitem{matson_methyl_1976}
G.~B. Matson {\em J.\,Chem.\,Phys.}, vol.~65, p.~4147, 1976.

\bibitem{matson_JCP_1977}
G.~B. Matson {\em J.\,Chem.\,Phys.}, vol.~67, p.~5152, 1977.

\bibitem{Werbelow_Advances_1977}
L.~G. Werbelow and D.~M. Grant, {\em Intramolecular Dipolar Relaxation in
  Multispin Systems}, vol.~9, p.~189.
\newblock Elsevier, 1977.

\bibitem{Muller_JMR_1987}
N.~Muller, G.~Bodenhausen, and R.~R. Ernst {\em J.\,Magn.\,Reson.}, vol.~75,
  p.~297, 1987.

\bibitem{Kay_JMR_1992}
L.~E. Kay and T.~Bull {\em J.\,Magn.\,Reson.}, vol.~99, p.~615, 1992.

\bibitem{Dumez_JCP_2015}
J.-N. Dumez, P.~Hakansson, S.~Mamone, B.~Meier, G.~Stevanato, J.~T.
  Hill-Cousins, S.~Singha~Roy, R.~C. Brown, G.~Pileio, and M.~H. Levitt {\em
  J.\,Chem.\,Phys.}, vol.~142, p.~044506, 2015.

\bibitem{LevittC3}
M.~Levitt. personal communication, 2019-02-02.

\bibitem{kumar_cross-correlations_2000}
A.~Kumar, R.~Christy Rani~Grace, and P.~K. Madhu {\em
  Prog.\,Nucl.\,Magn.\,Reson.}, vol.~37, p.~191, 2000.

\bibitem{goldman_interference_1984}
M.~Goldman {\em J.\,Magn.\,Reson.}, vol.~60, p.~437, 1984.

\bibitem{kowalewski_nuclear_2006}
J.~Kowalewski and L.~M{\"a}ler, {\em Nuclear Spin Relaxation in Liquids:
  Theory, Experiments, and Applications}.
\newblock Taylor \& Francis, 2006.

\bibitem{nicholas_nuclear_2010}
M.~P. Nicholas, E.~Eryilmaz, F.~Ferrage, D.~Cowburn, and R.~Ghose {\em
  Prog.\,Nucl.\,Magn.\,Reson.}, vol.~57, p.~111, 2010.

\bibitem{abragam_principles_1961}
A.~Abragam, {\em The Principles of Nuclear Magnetism}.
\newblock Oxford University Press, London, 1961.

\bibitem{lipari_model-free_1982}
G.~Lipari and A.~Szabo {\em J.\,Am.\,Chem.\,Soc.}, vol.~104, p.~4546, 1982.

\bibitem{Clore_JACS_1990}
M.~G. Clore, A.~Szabo, A.~Bax, K.~E. Lewis, P.~C. Driscoll, and A.~M.
  Bronenborn {\em J.\,Am.\,Chem.\,Soc.}, vol.~112, p.~4989, 1990.

\bibitem{Liao_JPC_2012}
X.~Liao, D.~Long, D.-W. Li, R.~Bruschweiler, and V.~Tugarinov {\em
  J.\,Phys.\,Chem.\,B}, vol.~116, p.~606, 2012.

\bibitem{Frueh_PNMR_2002}
D.~Frueh {\em Prog.\,Nucl.\,Magn.\,Reson.}, vol.~41, p.~305, 2002.

\bibitem{Tugarinov_JBNMR_2004}
V.~Tugarinov, C.~Scheurer, R.~Bruschweiler, and L.~E. Kay {\em
  J.\,Biomol.\,NMR}, vol.~30, p.~397, 2004.

\bibitem{Tugarinov_JACS_2007_bis}
V.~Tugarinov and L.~E. Kay {\em J.\,Am.\,Chem.\,Soc.}, vol.~129, p.~9514, 2007.

\bibitem{Batchelder_JACS_1983}
L.~S. Batchelder, C.~Niu, and D.~Torchia {\em J.\,Am.\,Chem.\,Soc.}, vol.~105,
  p.~2228, 1983.

\bibitem{Kaderavek_JCPC_2017}
P.~Kaderavek, L.~Strouk, S.~F. Cousin, C.~Charlier, G.~Bodenhausen,
  T.~Marquardsen, J.-M. Tyburn, P.-A. Bovier, F.~Engelke, W.~Maas, and
  F.~Ferrage {\em ChemPhysChem}, vol.~18, p.~2772, 2017.

\bibitem{Charlier_JACS_2013}
C.~Charlier, S.~N. Khan, T.~Marquardsen, P.~Pelupessy, V.~Reiss,
  D.~Sakellariou, G.~Bodenhausen, F.~Engelke, and F.~Ferrage {\em
  J.\,Am.\,Chem.\,Soc.}, vol.~135, p.~18665, 2013.

\end{thebibliography}


@PREAMBLE{
 "\providecommand{\noopsort}[1]{}" 
 # "\providecommand{\singleletter}[1]{#1}
}


@inbook{Werbelow_Advances_1977,
  title = "Intramolecular Dipolar Relaxation in Multispin Systems",
  booktitle = "Advances in Magnetic Resonance",
  editor = "J.S. Waugh",
  publisher = "Elsevier",
  author = "Werbelow, Lawrence G. and Grant, David M.",
  Volume = "9",
  year = "1977",
  pages = "189",
}

@article{Tugarinov_JACS_2007_bis,
  volume = "129",
  journal = "J.\,Am.\,Chem.\,Soc.",
  author = "Tugarinov, Vitali and Kay, Lewis E.",
  year = "2007",
  pages = "9514",
  }

@article{Sheppard_Progress_2010,
  volume = "56",
  journal = "Prog.\,Nucl.\,Magn.\,Reson.",
  author = "Sheppard, Devon and Sprangers, Remco and Tugarinov, Vitali",
  year = "2010",
  pages = "1",
  }

@article{Goldman_JMagRes_1984,
  volume = "60",
  journal = "J.\,Magn.\,Reson.",
  author = "Goldman, Maurice",
  year = "1984",
  pages = "437",
  }

@article{Shimizu_JCP_1964,
  volume = "40",
  journal = "J.\,Chem.\,Phys.",
  author = "Shimizu, Hiroshi",
  year = "1964",
  pages = "3357",
  }

@article{Wimperis_MolPhys_1989,
  volume = "66",
  journal = "Molec.\,Phys.",
  author = "Wimperis, Stephen and Bodenhausen, Geoffrey",
  year = "1989",
  pages = "897",
  }

@article{Kay_JMR_1992,
  volume = "99",
  journal = "J.\,Magn.\,Reson.",
  author = "Kay, Lewis E. and Bull, T.E.",
  year = "1992",
  pages = "615",
  }

@article{matson_JCP_1977,
  volume = "67",
  journal = "J.\,Chem.\,Phys.",
  author = "Matson, Gerald B.",
  year = "1977",
  pages = "5152",
  }

@article{Muller_JMR_1987,
  volume = "75",
  journal = "J.\,Magn.\,Reson.",
  author = "Muller, Norbert and Bodenhausen, Geoffrey and Ernst, Richard R.",
  year = "1987",
  pages = "297",
}

@article{Liao_JPC_2012,
  volume = "116",
  journal = "J.\,Phys.\,Chem.\,B",
  author = "Liao, Xinli and Long, Dong and Li, Da-Wei and Bruschweiler, Rafel and Tugarinov, Vitali",
  year = "2012",
  pages = "606",
}

@article{Frueh_PNMR_2002,
  volume = "41",
  journal = "Prog.\,Nucl.\,Magn.\,Reson.",
  author = "Frueh, Dominique",
  year = "2002",
  pages = "305",
}

@misc{LevittC3,
  author = "Levitt, Malcolm",
  year = "2019-02-02",
  howpublished = "personal communication"
}


@article{Charlier_JACS_2013,
  volume = "135",
  journal = "J.\,Am.\,Chem.\,Soc.",
  author = "Charlier, Cyril and Khan, Shahid Nawaz and Marquardsen, Thorsten and Pelupessy, Philippe and Reiss, Volker and Sakellariou, Dimitris and Bodenhausen, Geoffrey and Engelke, Frank and Ferrage, Fabien",
  year = "2013",
  pages = "18665",
}


@article{Kaderavek_JCPC_2017,
  volume = "18",
  journal = "ChemPhysChem",
  author = "Kaderavek, Pavel and Strouk, Léonard and Cousin, Samuel F. And Charlier, Cyril and Bodenhausen, Geoffrey and Marquardsen, Thorsten and Tyburn, Jean-Max and Bovier, Pierre-Alain and Engelke, Frank and Maas, Werner and Ferrage, Fabien",
  year = "2017",
  pages = "2772",
}


@article{Batchelder_JACS_1983,
  volume = "105",
  journal = "J.\,Am.\,Chem.\,Soc.",
  author = "Batchelder, Lynne S. And Niu, C.H. and Torchia, D.A",
  year = "1983",
  pages = "2228",
}


@article{Tugarinov_JBNMR_2004,
  volume = "30",
  journal = "J.\,Biomol.\,NMR",
  author = "Tugarinov, Vitali and Scheurer, Christoph and Bruschweiler, Rafael and Kay, Lewis E.",
  year = "2004",
  pages = "397",
}

@article{Dumez_JCP_2015,
  volume = "142",
  journal = "J.\,Chem.\,Phys.",
  author = "Dumez, Jean-Nicolas and Hakansson, Par and Mamone, Salvatore and Meier, Benno and Stevanato, Gabriele and Hill-Cousins, Joseph T. And Singha Roy, Soumya and Brown, Richard C.D. and Pileio, Giuseppe and Levitt, Malcolm H.",
  year = "2015",
  pages = "044506",
}


@article{Blumich_ACIE_2017,
  volume = "57",
  journal = "Angew. Chem. Int. Ed.",
  author = "Blumich, Bernhard and Singh, Kawarpal",
  year = "2017",
  pages = "6996",
}


@article{Gan_JMR_2017,
  volume = "284",
  journal = "J.\,Magn.\,Reson.",
  author = "Gan, Zhehong and Hung, Ivan and Wang, Xiaoling and Paulino, Joana and Gang, Wu and Litvak, Ilya M. and Gor'kov, Peter L. and Grey, William W. and Lendi, Pietro and Schiano, Jeffrey L. and Bird, Mark D. and Dixon, Iain R. and Toth, Jack and Boebinger, Gregory and Cross, Timothy A.",
  year = "2017",
  pages = "125",
}



@article{Helmus_JbiolNMR_2013,
  volume = "55",
  journal = "J.\,Biomol.\,NMR",
  author = "Helmus, Jonathan J. and Jaroniec, Christopher P.",
  year = "2013",
  pages = "355-367",
}

@article{frederick_conformational_2007,
  volume = "448",
  journal = "Nature",
  author = "Frederick, Kendra King and Marlow, Michael S. and Valentine, Kathleen G. and Wand, A. Joshua",
  year = "2007",
  pages = "325",
}

@article{matson_methyl_1976,
  volume = "65",
  journal = "J.\,Chem.\,Phys.",
  author = "Matson, Gerald B.",
  year = "1976",
  pages = "4147",
  }

@article{tugarinov_line_2004,
  volume = "126",
  journal = "J.\,Am.\,Chem.\,Soc.",
  author = "Tugarinov, Vitali and Sprangers, Remco and Kay, Lewis E.",
  year = "2004",
  pages = "4921",
}

@article{kuprov_bloch-redfield-wangsness_2007,
  volume = "184",
  journal = "J.\,Magn.\,Reson.",
  author = "Kuprov, Ilya and Wagner-Rundell, Nicola and Hore, P.J.",
  year = "2007",
  pages = "196",
}

@software{wolfram_research_inc._mathematica_2016,
  location = "Champaign, Illinois",
  title = "Mathematica",
  version = "10.4",
  timestamp = "2016-06-06T16:52:06Z",
  publisher = "Wolfram Research, Inc.",
  author ="Wolfram Research, Inc.",
  year = "2016",
}

@article{cousin_high-resolution_2016,
  journal = "Phys.\,Chem.\,Chem.\,Phys.",
  author = "Cousin, Samuel F. and Charlier, Cyril and Kade{\v r}{\'a}vek, Pavel and Marquardsen, Thorsten and Tyburn, Jean-Max and Bovier, Pierre-Alain and Ulzega, Simone and Speck, Thomas and Wilhelm, Dirk and Engelke, Frank and Maas, Werner and Sakellariou, Dimitrios and Bodenhausen, Geoffrey and Pelupessy, Philippe and Ferrage, Fabien",
  year = "2016",
  volume = "18",
  pages = "33187",
}

@article{Werbelow_JCP_1975,
  volume = "63",
  journal = "J.\,Chem.\,Phys.",
  author = "Werbelow, Lawrence G. and Grant, David M.",
  year = "1975",
  pages = "544",
}

@article{Clore_JACS_1990,
  volume = "112",
  journal = "J.\,Am.\,Chem.\,Soc.",
  author = "Clore, Marius G. and Szabo, Attila and Bax, Ad and Lewis, Kay E. and Driscoll, Paul C. and Bronenborn, Angela M.",
  year = "1990",
  pages = "4989",
}

@article{lipari_model-free_1982,
  volume = "104",
  journal = "J.\,Am.\,Chem.\,Soc.",
  author = "Lipari, Giovanni and Szabo, Attila",
  year = "1982",
  pages = "4546",
}

@article{Bengs_MagnResonChecm_2017,
  volume = "56",
  journal = "Magn.\,Reson.\,Chem.",
  author = "Bengs, Christian and Levitt, Malcolm H.",
  year = "2018",
  pages = "374",
}

@book{abragam_principles_1961,
  title = "The Principles of Nuclear Magnetism",
  publisher = "Oxford University Press, London",
  author = "Abragam, Anatole",
  year = "1961",
}

@article{nicholas_nuclear_2010,
  volume = "57",
  journal = "Prog.\,Nucl.\,Magn.\,Reson.",
  author = "Nicholas, Matthew P. and Eryilmaz, Ertan and Ferrage, Fabien and Cowburn, David and Ghose, Ranajeet",
  year = "2010",
  pages = "111",
}

@book{kowalewski_nuclear_2006,
  title = "Nuclear Spin Relaxation in Liquids: Theory, Experiments, and Applications",
  publisher = "Taylor \& Francis",
  author = "Kowalewski, J{\'o}zef and M{\"a}ler, Lena",
  year = "2006",
}

@article{goldman_interference_1984,
  volume = "60",
  journal = "J.\,Magn.\,Reson.",
  author = "Goldman, Maurice",
  year = "1984",
  pages = "437",
}

@article{kumar_cross-correlations_2000,
  volume = "37",
  journal = "Prog.\,Nucl.\,Magn.\,Reson.",
  author = "Kumar, Anil and Christy Rani Grace, R. and Madhu, P. K.",
  year = "2000",
  pages = "191",
}

@article{tugarinov_JACS_2007,
  volume = "129",
  journal = "J.\,Am.\,Chem.\,Soc.",
  author = "Tugarinov, Vitali and Sprangers, Remco and Kay, Lewis E.",
  year = "2007",
  pages = "1743",
}

@article{cousin_recovering_2016,
  journal = "Angew.\,Chem.\,Int.\,Ed.",
  author = "Cousin, Samuel F. and Kade{\v r}{\'a}vek, Pavel and Haddou, Baptiste and Charlier, Cyril and Marquardsen, Thorsten and Tyburn, Jean-Max and Bovier, Pierre-Alain and Engelke, Frank and Maas, Werner and Bodenhausen, Geoffrey and Pelupessy, Philippe and Ferrage, Fabien",
  year = "2016",
  volume = "55",
  pages = "9886",
}

@article{rosenzweig_bringing_2014,
  volume = "83",
  journal = "Annu.\,Rev.\,Biochem.",
  author = "Rosenzweig, Rina and Kay, Lewis E.",
  year = "2014",
  pages = "291",
}

@inbook{clore_chapter_2012,
  title = "Isotope-Labelling of Methyl Groups for NMR Studies of Large Proteins",
  booktitle = "RSC Biomolecular Sciences",
  publisher = "Royal Society of Chemistry",
  author = "Plevin, Michael J. and Boisbouvier, J{\'e}r{\^o}me",
  chapter = "1",
  year = "2012",
  pages = "1--24",
}

@article{ollerenshaw_comparison_2005,
  volume = "33",
  journal = "J.\,Biomol.\,NMR",
  author = "Ollerenshaw, Jason E. and Tugarinov, Vitali and Skrynnikov, Nikolai R. and Kay, Lewis E.",
  year = "2005",
  pages = "25",
}

@article{tugarinov_cross-correlated_2003,
  volume = "125",
  journal = "J.\,Am.\,Chem.\,Soc.",
  author = "Tugarinov, Vitali and Hwang, Peter M. and Ollerenshaw, Jason E. and Kay, Lewis E.",
  year = "2003",
  pages = "10420",
}


@article{matson_methyl_1977,
  volume = "67",
  journal = "J.\,Chem.\,Phys.",
  author = "Matson, Gerald B.",
  year = "1977",
  pages = "5152",
}

@article{mas_SciAdv_2018,
  volume = "4",
  journal = "Sci.\,Adv.",
  author = "Mas, Guillaume and Guan, Jia-Ying and Crublet, Elodie and Colas Debled, Elisa and Moriscot, Christine and Gans, Pierre and Schoehn, Guy and Macek, Paul and Boisbouvier, Jerome",
  year = "2018",
  pages = "1",
}

@article{Cousin_JACS_2018,
  volume = "140",
  journal = "J.\,Am.\,Chem.\,Soc.",
  author = "Cousin, Samuel F. and Kaderavek, Pavel and Bolik-Coulon, Nicolas and Gu, Yina and Charlier, Cyril and Carlier, Ludovic and Bruschweiler-Li, Lei and Marquardsen, Thorsten and Tyburn, Jean-Max and Bruschweiler, Rafael and Ferrage, Fabien",
  year = "2018",
  pages = "13456",
}

@article{nicholson_dynamics_1992,
  volume = "31",
  journal = "Biochemistry",
  author = "Nicholson, Linda K. and Kay, Lewis E. and Baldisseri, Donna M. and Arango, Julian and Young, Paul E. and Bax, Ad and Torchia, Dennis A.",
  year = "1992",
  pages = "5253",
}

@article{kasinath_AngevChem_2013,
  volume = "54",
  journal = "Angew.\,Chem.\,Int.\,Ed.",
  author = "Kasinath, Vignesh and Fu, Yinan and Sharp, Kim A. and Wand, A. Joshua",
  year = "2014",
  pages = "102",
}

@article{DuBay_AccChemRes_2015,
  volume = "48",
  journal = "Acc.\,Chem.\,Res.",
  author = "DuBay, Kateri H. and Bowman, Gregory R. and Geissler, Phillip L.",
  year = "2015",
  pages = "1098",
}

@article{hajduk_nmr-based_2000,
  volume = "122",
  journal = "J.\,Am.\,Chem.\,Soc.",
  author = "Hajduk, Philip J. and Augeri, David J. and Mack, Jamey and Mendoza, Renaldo and Yang, Jianguo and Betz, Stephen F. and Fesik, Stephen W.",
  year = "2000",
  pages = "7898",
}

@article{tugarinov_quantitative_2005,
  volume = "44",
  journal = "Biochemistry",
  author = "Tugarinov, Vitali and Kay, Lewis E.",
  year = "2005",
  pages = "15970",
}

@article{frederick_conformational_2007-1,
  volume = "448",
  journal = "Nature",
  author = "Frederick, Kendra King and Marlow, Michael S. and Valentine, Kathleen G. and Wand, A. Joshua",
  year = "2007",
  pages = "325",
}

@article{mas_specific_2013,
  volume = "57",
  journal = "J.\,Biomol.\,NMR",
  author = "Mas, Guillaume and Crublet, Elodie and Hamelin, Olivier and Gans, Pierre and Boisbouvier, J{\'e}r{\^o}me",
  year = "2013",
  pages = "251",
}

@article{fiaux_nmr_2002,
  volume = "418",
  journal = "Nature",
  author = "Fiaux, Jocelyne and Bertelsen, Eric B. and Horwich, Arthur L. and Wuthrich, Kurt",
  year = "2002",
  pages = "207",
}

@article{tugarinov_isotope_2006,
  volume = "1",
  journal = "Nat.\,Protoc.",
  author = "Tugarinov, Vitali and Kanelis, Voula and Kay, Lewis E",
  year = "2006",
  pages = "749",
}

@article{korzhnev_probing_2004,
  volume = "126",
  journal = "J.\,Am.\,Chem.\,Soc.",
  author = "Korzhnev, Dmitry M. and Kloiber, Karin and Kanelis, Voula and Tugarinov, Vitali and Kay, Lewis E.",
  year = "2004",
  pages = "3964",
}

@article{Tugarinov_fourdimensional_2002,
  volume = "124",
  journal = "J.\,Am.\,Chem.\,Soc.",
  author = "Tugarinov, Vitali and Muhandiram, Ranjith and Ayed, Ayeda and Kay, Lewis E.",
  year = "2002",
  pages = "10025",
}

@article{Pervushin_attenuated_1997,
  volume = "94",
  journal = "Proc.\,Natl.\,Acad.\,Sci.\,USA",
  author = "Pervushin, Konstantin and Riek, Roland and Wider, Gerhard and Wuthrich, Kurt",
  year = "1997",
  pages = "12366",
}

@article{medek_multiple-quantum_1995,
  volume = "1117",
  journal = "J.\,Am.\,Chem.\,Soc.",
  author = "Medek, Ales and Harwood, John S. and Frydman, Lucio",
  year = "1995",
  pages = "12779",
}


@article{Frederick_Nature_2007,
  volume = "448",
  journal = "Nature",
  author = "Frederick, Kendra King and Marlow, Michael S. and Valentine, Kathleen G. and Wand, Joshua",
  year = "2007",
  pages = "325",
}


@article{nowakowski_applications_2015,
  volume = "90-91",
  journal = "Prog.\,Nucl.\,Magn.\,Reson.",
  author = "Nowakowski, Micha{\l} and Saxena, Saurabh and Stanek, Jan and {\.Z}erko, Szymon and Ko{\'z}mi{\'n}ski, Wiktor",
  year = "2015",
  pages = "49",
}

@article{schanda_speeding_2006,
  volume = "128",
  journal = "J.\,Am.\,Chem.\,Soc.",
  author = "Schanda, Paul and Van Melckebeke, H{\'e}l{\`e}ne and Brutscher, Bernhard",
  year = "2006",
  pages = "9042",
}

@article{Charlier_ChemSocRev_2016,
  volume = "45",
  journal = "Chem.\,Soc.\,Rev.",
  author = "Charlier, Cyril and Cousin, Samuel and Ferrage, Fabien",
  year = "2016",
  pages = "2410",
}

@article{kateb_slow_2006,
  volume = "45",
  journal = "Biochemistry",
  author = "Kateb, Fatiha and Abergel, Daniel and Blouquit, Yves and Duchambon, Patricia and Craescu, Constantin T. and Bodenhausen, Geoffrey",
  year = "2006",
  pages = "15011",
}

@article{ferrage_protein_2006,
  volume = "128",
  journal = "J.\,Am.\,Chem.\,Soc.",
  author = "Ferrage, Fabien and Pelupessy, Philippe and Cowburn, David and Bodenhausen, Geoffrey",
  year = "2006",
  pages = "11072",
}

@article{chevelkov_comparison_2010,
  volume = "132",
  journal = "J.\,Am.\,Chem.\,Soc.",
  author = "Chevelkov, Veniamin and Xue, Yi and Linser, Rasmus and Skrynnikov, Nikolai R. and Reif, Bernd",
  year = "2010",
  pages = "5015",
}

@article{schanda_quantitative_2010,
  volume = "132",
  journal = "J.\,Am.\,Chem.\,Soc.",
  author = "Schanda, Paul and Meier, Beat H. and Ernst, Matthias",
  year = "2010",
  pages = "15957",
}

@article{ramboarina_probing_2008,
  volume = "130",
  journal = "J.\,Am.\,Chem.\,Soc.",
  author = "Ramboarina, St{\'e}phanie and Redfield, Christina",
  year = "2008",
  pages = "15318",
}

@article{mandel_backbone_1995,
  volume = "246",
  journal = "J.\,Mol.\,Biol.",
  author = "Mandel, Arthur M. and Akke, Mikael and Palmer, III, Arthur G.",
  year = "1995",
  pages = "144",
}

@article{farrow_backbone_1994,
  volume = "33",
  journal = "Biochemistry",
  author = "Farrow, Neil A. and Muhandiram, Ranjith and Singer, Alex U. and Pascal, Steven M. and Kay, Cyril M. and Gish, Gerry and Shoelson, Steven E. and Pawson, Tony and Forman-Kay, Julie D. and Kay, Lewis E.",
  year = "1994",
  pages = "5984",
}

@article{Ardenkjaer_Angewandte_2015,
  volume = "54",
  journal = "Angew.\,Chem.\,Int.\,Ed.",
  author = "Ardenkjaer-Larsen, Jan-Henrik and Boebinger, Gregory S. and Comment, Arnaud A. and Duckett, Simon and Edison, Arthur S. and Engelke, Frank and Griesinger, Christian and Griffin, Robert G. and Hilty, Christian and Maeda, Hidaeki and Parigi, Giacomo and Prisner, Thomas and Ravera, Enrico and van Bentum, Jan and Vega, Shimon and Webb, Andrew and Luchinat, Claudio and Schwalbe, Harald and Frydman, Lucio",
  year = "2015",
  pages = "9162",
}


@article{kay_backbone_1989,
  volume = "28",
  journal = "Biochemistry",
  author = "Kay, Lewis E. and Torchia, Dennis A. and Bax, Ad",
  year = "1989",
  pages = "8972",
}


@article{Mittermaier_science_2006,
  volume = "312",
  journal = "Science",
  author = "Mittermaier, Anthony and Kay, Lewis E.",
  year = "2006",
  pages = "224",
}

@article{Palmer_ChemRev_2004,
  volume = "104",
  journal = "Chem.\,Rev.",
  author = "Palmer, Arthur G",
  year = "2004",
  pages = "3640",
}

@book{Cavanagh_ProteinNMR_2007,
  title = "Protein NMR Spectroscopy: Principles and Practice",
  publisher = "Elsevier Academic Press",
  author = "Cavanagh, John and Fairbrother, Wayne J. and Palmer, Arthur G. and Rance, Mark and Skelton, Nicholas J.",
  year = "2007"
}


@preamble{ "\ifdefined\DeclarePrefChars\DeclarePrefChars{''-}\else\fi " }
